# Issues on drawing the State Transition Diagram for arbitrary Cellular Automata

**Sudhakar Sahoo, Pabitra Pal Choudhury**
Applied Statistics Unit, Indian Statistical Institute, Kolkata, 700108, INDIA
Email: sudhakar.sahoo@gmail.com & pabitrapalchoudhury@gmail.com

**Abstract:** This paper proposes several algorithms and their Cellular Automata Machine (CAM) for drawing the State Transition Diagram (STD) of an arbitrary Cellular Automata (CA) Rule $x$ (any neighborhood, uniform/ hybrid and null/ periodic boundary) and length of the CA n. It also discusses the novelty, hardware cost and the complexities of these algorithms.

## 1. Introduction

J. von Neumann [1] framed Cellular Automata (CA) as a cellular space capable of self-reproduction. Since then many researchers have taken interest in the study of CA for modeling the behavior of complex systems. Wolfram [2] studied one-dimensional CA with the help of Space-Time diagrams as well as using State Transition Diagrams (STD's).

STD of CA rules both in null and periodic boundary finds its application in various inter disciplinary fields. In terms of dynamical systems, the STD is a representation of the phase space of a CA. Various other names are used for STD's, for example Basins of Attraction, flow graphs and networks of attraction [4, 5]. In recent years a group of researchers from Cellular Automata Research Lab, [8-11] used Basins of Attractors to solve many real life problems like Pattern Classification, Cryptology, Data compression, Bio informatics, Density Classification etc. using both null and periodic boundary CA. In [12] using 2D null boundary CA linear rules and its matrices we had obtained some fundamental image processing operations like zooming in, zooming out, multiple copy of the image etc., for certain symmetric images. Besides their potential applications, the theoretical study of STD both in null and periodic has mathematical interest on their own. In [4] Wuensche and Lesser provides the motivation and presents two different algorithms: exhaustive testing and direct reverse algorithms for drawing STDs of CA rules in periodic boundary conditions. Using STD's Das et al. [6, 7] characterizes reachable/non-reachable CA states for null boundary 1D CA rules and classified all the CA rules with a target to synthesize the Reversible CA. In [13] Kang et al. extended this work and provides two decision algorithms for classification of reachable/non-reachable state and group/non-group CA based on periodic boundary condition. Again for periodic boundary CA, later publications by the same author as [4] provided exhaustive algorithms and the efficient direct reverse algorithms for CA with mixed rules and/or arbitrary connections, for the more general Random Boolean Network (RBN) [5], discrete (multi-value) networks, and for random maps, implemented in DDLab and described in its manual [17].

One advantage of using direct reverse algorithms is that it is much easier to draw partial STDs (e.g. only a single attractor basin). Because the complete STD contains all the $2^n$ states, therefore the time required to draw complete STD containing all the states cannot be less than $2^n$. This proves that the problem is in EXPTIME and thus intractable for large n [14, 15]. So for an arbitrary CA rule both in null and periodic boundary drawing the complete STD of an n-length CA is possible only for small values of n and observing these complete STD's and with the help of other mathematical tools like deviant states, matrix theory etc. [18] one can predict some of the characteristics of CA (reachable/non-reachable states, Predecessors, Self-loops, Point attractors etc.) for sufficiently large values of n. Further, when n is very large, on observing even the partial STD's one can also get some of the characteristics of the corresponding CA rule for the purpose of analysis.

In this paper our aim is to obtain different algorithms and their Cellular Automata Machine (CAM) architectures [16] to draw the complete STD of arbitrary CA rules of length n both in Null (NB) and Periodic (PB) Boundary conditions. First algorithm is very similar to the existing exhaustive testing algorithm [4] where, the nodes are arranged in a table in ascending order of their decimal value of the states, starting from the state 0 to state $2^n - 1$ and the corresponding CAM architecture for this algorithm is proposed. The ascending order of the state arrangement of this algorithm produces interesting bit patterns where the novelty of our methodology lies. From this algorithm another efficient algorithm and its CAM architecture are designed. So our exposition centers on the methodologies where by various characteristics of STD like reachable and non-reachable states, predecessors, successors, self-loop etc. are brought out. Since the problem is in EXPTIME, so for drawing the complete STD, direct reverse algorithm which do not require a table of exhaustive pairs is efficient (as mentioned in [5]) than the exhaustive testing algorithm in the sense that, they posses the same order of complexity $O(2^n)$ but the associated constants would be different. It is to be noted that we have analyzed the existing exhaustive testing algorithm as well as our two newly introduced algorithms with respect to time and space complexity. Also notice that the CAM for our algorithm has been designed where as CAM for the direct reverse algorithm does not seem to exist. For our CAM design we have computed the time, space and the wiring costs for both the newly introduced algorithms. Similar algorithms and their CAM for the more general Random Boolean Network (RBN) [5], discrete (multi-value) networks, and for random maps, implemented in DDLab and described in its manual [17] can also be obtained which is beyond the scope of this paper and can be dealt in a separate paper. The organization of the current paper is as follows:

Basic concepts and various terminology of this study are defined in Section 2. Section 3 presents the algorithms and their CAM architecture for drawing the STD of null and periodic boundary uniform CA with a given rule $x$ and length of the CA n. Descriptions are also given for CA with mixed rules. Also the correctness, novelties, hardware and software complexities etc. of these algorithms are discussed in this section. Finally a conclusion is drawn in Section 4.

## 2 Terminology and notation pertaining to one-dimensional Cellular Automata

In this paper, we shall restrict ourselves to the study of a deterministic one-dimensional, binary cellular automaton (CA) of *n* cells (i.e. *n* bits) $x_1$, $x_2 \ldots x_n$. The rule-table order and decimal rule numbering used here is by Wolfram's convention [3], which also defines Periodic (PB) and Null (NB) boundary conditions. Quick reviews of some of the related CA definitions are follows.

The ***global state*** or simply *state* of a CA at any time-instant *t* is represented as a vector $X^t = (x_1^t, x_2^t \ldots x_n^t)$ where $x_i^t$ denotes the bit in the $i^{th}$ cell $x_i$ at time-instant t. However, instead of expressing a state as a bit-string, it can also be represented by the decimal equivalent of the *n*-bit string with $x_1$ as the Most Significant Bit; e.g. for a 4-bit CA, the state *1011* may be referred to as state 11 $(=1\times2^0 + 1\times2^1 + 0\times2^2 + 1\times2^3)$. It can be noted that the conversion of the states to its decimal notation is a shorthand representation and does not impact on the complexity of STD drawing algorithms and there would not be any problem on leaving the states in binary notation.

The bit in the $i^{th}$ cell at the "next" time-instant $t$+1 is given by a ***local mapping*** denoted by $f^i$, say, which takes as its argument a vector of the bits (in proper order) at time-instant t in the cells of a certain pre-defined ***neighborhood*** (of size *p*, say) of the $i^{th}$ cell. Thus, the size of the neighborhood is taken to be the same for each cell and may also be called the 'number of variables' (which $f^i$ takes as inputs). For our purpose, we shall be mostly interested in ***elementary***

***CA*** to be one-dimensional binary CA with a symmetrical neighborhood of size $p$ = 3 for each cell so that $x_i^{t+1} = f^{\,i}(x_{i-1}^t, x_i^t, x_{i+1}^t)$ , $i$ = 2,3,……., $n$-1.

**Null boundary (*NB*):** The left neighbor of $x_1$ and the right neighbor of $x_n$ are taken as 0 each.

**Periodic boundary (*PB*):** $x_n$ is taken as the left neighbor of $x_1$ and $x_1$ as the right neighbor of $x_n$.

A CA may be represented as a string of the rules applied to the cells in proper order, along with a specification of the boundary conditions. e.g. <103, 234, 90, 0>NB refers to the CA ($x_1$, $x_2$, $x_3$, $x_4$) where $x_1^{t+1} = f_{103}(0, x_1^t, x_2^t)$; $x_2^{t+1} = f_{234}(x_1^t, x_2^t, x_3^t)$; $x_3^{t+1} = f_{90}(x_2^t, x_3^t, x_4^t)$; $x_4^{t+1} = f_0(x_3^t, x_4^t, 0)$. If the "present state" of an $n$-bit CA (at time $t$) is $X^t$, its "next state" (at time $t+1$), denoted by $X^{t+1}$, is in general given by the ***global mapping*** $F(X^t) = ( f^1(lb^t, x_1^t, x_2^t), f^2(x_1^t, x_2^t, x_3^t),\ldots, f^n(x_{n-1}^t, x_n^t, rb^t) )$, where $lb$ and $rb$ denote respectively the left boundary of $x_1$ and right boundary of $x_n$. If the same Boolean function (rule) determines the "next" bit in each cell of a CA, the CA will be called a **Uniform Cellular Automaton (UCA)**, otherwise it will be called a **Hybrid Cellular Automaton (HCA)**, e.g.<135, 135, 135, 135>PB is a UCA, <0, 60, 72, 72>NB is a HCA.

For a UCA, the Boolean function applied to each cell will be called the rule of the CA. So for a UCA, we can simply denote it as $f_R$. Henceforth, we shall use the following notation, UCA*n*NB, UCA*n*PB, HCA*n*NB and HCA*n*PB for an *n*-bit CA.

**Cellular Automata Machine (CAM):** As mentioned by Tommaso Toffoli and Norman Margolis [16], the structure of a CA is ideally suited for realization on a machine having a high degree of parallelism with local and uniform interconnections. With an appropriate architecture one can achieve in the simulation of CA a performance at least several orders of magnitude greater than that of a conventional computer, for a comparable cost. One can view CAM much like an appliance-in, which a few rows of buttons directly control, in an interactive manner, a number of functions and options. The CAM(s) designed in this paper can be viewed in the same way as in [16].

## 3. State Transition Diagram for CA rules

The evolution of a CA can be completely described by a diagram in which each state is connected to its successor by a properly directed line-segment. This diagram is called the State Transition Diagram (abbreviated as STD) of the CA. In other words, the STD of a CA is essentially a directed graph where each ***node*** represents one of the states of the CA and the ***edges*** signify transitions from one state to another. Wuensche et al. [4, 5, 17] provides two different algorithms: exhaustive testing and the direct reverse algorithms for drawing STDs of arbitrary CA rules. Although their algorithms are for periodic boundary conditions but it can also be modified for null boundary conditions.

The approach of the most obvious exhaustive testing algorithms is to create a table where every state in state-space is paired with its successor (the successor of X is the state Y such that F(X) = Y). States missing from the successor list are leaves in the STD, states that appear once or more give predecessors [19]. Where as the approach of the direct reverse algorithms is to "work backwards", making use of the algorithm for finding the pre images of a CA state (the pre images of X are the states Y such that F(Y) = X). Start at a state on an attractor cycle, find its pre images, and find each pre image's pre images, and so on. The advantage of this approach is that it is much easier to draw partial STDs (e.g. only a single attractor basin) and it is more efficient than the exhaustive testing algorithm.

As discussed earlier, for an n-length CA the complete STD contains all the $2^n$ states, therefore the time required to draw the complete STD containing all the states cannot be less than $2^n$. This proves that the problem is in EXPTIME and intractable [14, 15]. Thus for drawing the complete STD, direct reverse algorithm is efficient than the exhaustive testing algorithm in the sense that, they posses the same order of complexity $O(2^n)$ but the associated constants would be different. Because of exponential complexity, drawing the complete STD for an arbitrary CA rule of length-n is only possible for small values of n. On observing these complete STD's and with the help of other mathematical tools like deviant states, matrix theory etc. [18] one can predict some of the characteristics (reachable/non-reachable states, Predecessors, Self-loops, Point attractors etc.) for sufficiently large values of n. Further, when n is very large, on observing even the partial STD's one can get some of the characteristics of the corresponding CA rule for the purpose of analysis.

In this section we have developed two new algorithms and their Cellular Automata Machine's (CAM) [16] to draw the complete STD of arbitrary CA rules of length n both for Null (NB) and Periodic (PB) Boundary conditions. First algorithm is very similar to the existing exhaustive testing algorithm as reported in [4] where, the nodes are arranged in a table in ascending order of their decimal value of the states, starting from the state 0 to state $2^n - 1$ and the corresponding CAM architecture for this algorithm is proposed. The ascending order of the state arrangement produces interesting bit patterns where the novelty lies. From this algorithm another efficient algorithm and its CAM architecture are designed, which are more efficient in time, space and with respect to the wiring cost than the existing exhaustive testing algorithm for CA with mixed rules and/or arbitrary connections. Further, we have not compared the efficiency of our algorithms with the existing direct reverse algorithm because it seems that just like our way of analysis (in the order notation sense) and our proposed CAM design, no analysis of time and space complexities and the CAM architectures are available for the direct reverse algorithm [5].

The exhaustive testing algorithm can be demonstrated using the following example, which is the basis for our design.

**Example 3.1**
Suppose we have a Rule 52 UCA4NB. The truth table of this rule is shown in table 1.

| Dec. Value | X | Y | Z | $f_{52}$ |
|---|---|---|---|---|
| 0 | 0 | 0 | 0 | 0 |
| 1 | 0 | 0 | 1 | 0 |
| 2 | 0 | 1 | 0 | 1 |
| 3 | 0 | 1 | 1 | 0 |
| 4 | 1 | 0 | 0 | 1 |
| 5 | 1 | 0 | 1 | 1 |
| 6 | 1 | 1 | 0 | 0 |
| 7 | 1 | 1 | 1 | 0 |

[Table-1: Truth table for rule 52 in 3 variable Boolean function]

The process of exhaustive testing algorithm to draw the complete STD for Rule 52 UCA4NB is shown in table 2. This process has been translated into a software program by which the STD of Rule 52 can be drawn as shown in fig-1. If at an instant *t*, the state of the CA is **11** $\equiv$ 1011, that at *t*+1 will be ($f_{52}$(0, 1, 0), $f_{52}$ (1, 0, 1), $f_{52}$ (0, 1, 1), $f_{52}$ (1, 1, 0)) $\equiv$ ($f_{52}$(2), $f_{52}$ (5), $f_{52}$ (3), $f_{52}$ (6)) $\equiv$ (1, 1, 0, 0)) $\equiv$ **1100** $\equiv$ **12**. This transition is represented, by drawing an ***arrow*** from state **11** to its ***successor*** state **12**. Similarly, the successor of state **6** is **1** that of **15** is **0** and so on.

| States | $x_1^t$ | $x_2^t$ | $x_3^t$ | $x_4^t$ | $f_{52}$ | $x_1^{t+1}$ | $x_2^{t+1}$ | $x_3^{t+1}$ | $x_4^{t+1}$ | $\equiv$ | $x_1^{t+1}$ | $x_2^{t+1}$ | $x_3^{t+1}$ | $x_4^{t+1}$ | Successors |
|---|---|---|---|---|---|---|---|---|---|---|---|---|---|---|---|
| 0 | 0 | 0 | 0 | 0 | $\rightarrow$ | $f_{52}(0)$ | $f_{52}(0)$ | $f_{52}(0)$ | $f_{52}(0)$ | $\rightarrow$ | 0 | 0 | 0 | 0 | 0 |
| 1 | 0 | 0 | 0 | 1 | $\rightarrow$ | $f_{52}(0)$ | $f_{52}(0)$ | $f_{52}(1)$ | $f_{52}(2)$ | $\rightarrow$ | 0 | 0 | 0 | 1 | 1 |
| 2 | 0 | 0 | 1 | 0 | $\rightarrow$ | $f_{52}(0)$ | $f_{52}(1)$ | $f_{52}(2)$ | $f_{52}(4)$ | $\rightarrow$ | 0 | 0 | 1 | 1 | 3 |
| 3 | 0 | 0 | 1 | 1 | → | $f_{52}(0)$ | $f_{52}(1)$ | $f_{52}(3)$ | $f_{52}(6)$ | → | 0 | 0 | 0 | 0 | 0 |
| 4 | 0 | 1 | 0 | 0 | → | $f_{52}(1)$ | $f_{52}(2)$ | $f_{52}(4)$ | $f_{52}(0)$ | → | 0 | 1 | 1 | 0 | 6 |
| 5 | 0 | 1 | 0 | 1 | → | $f_{52}(1)$ | $f_{52}(2)$ | $f_{52}(5)$ | $f_{52}(2)$ | → | 0 | 1 | 1 | 1 | 7 |
| 6 | 0 | 1 | 1 | 0 | → | $f_{52}(1)$ | $f_{52}(3)$ | $f_{52}(6)$ | $f_{52}(4)$ | → | 0 | 0 | 0 | 1 | 1 |
| 7 | 0 | 1 | 1 | 1 | → | $f_{52}(1)$ | $f_{52}(3)$ | $f_{52}(7)$ | $f_{52}(6)$ | → | 0 | 0 | 0 | 0 | 0 |
| 8 | 1 | 0 | 0 | 0 | → | $f_{52}(2)$ | $f_{52}(4)$ | $f_{52}(0)$ | $f_{52}(0)$ | → | 1 | 1 | 0 | 0 | 12 |
| 9 | 1 | 0 | 0 | 1 | → | $f_{52}(2)$ | $f_{52}(4)$ | $f_{52}(1)$ | $f_{52}(2)$ | → | 1 | 1 | 0 | 1 | 13 |
| 10 | 1 | 0 | 1 | 0 | → | $f_{52}(2)$ | $f_{52}(5)$ | $f_{52}(2)$ | $f_{52}(4)$ | → | 1 | 1 | 1 | 1 | 15 |
| 11 | 1 | 0 | 1 | 1 | → | $f_{52}(2)$ | $f_{52}(5)$ | $f_{52}(3)$ | $f_{52}(6)$ | → | 1 | 1 | 0 | 0 | 12 |
| 12 | 1 | 1 | 0 | 0 | → | $f_{52}(3)$ | $f_{52}(6)$ | $f_{52}(4)$ | $f_{52}(0)$ | → | 0 | 0 | 1 | 0 | 2 |
| 13 | 1 | 1 | 0 | 1 | → | $f_{52}(3)$ | $f_{52}(6)$ | $f_{52}(5)$ | $f_{52}(2)$ | → | 0 | 0 | 1 | 1 | 3 |
| 14 | 1 | 1 | 1 | 0 | → | $f_{52}(3)$ | $f_{52}(7)$ | $f_{52}(6)$ | $f_{52}(4)$ | → | 0 | 0 | 0 | 1 | 1 |
| 15 | 1 | 1 | 1 | 1 | → | $f_{52}(3)$ | $f_{52}(7)$ | $f_{52}(7)$ | $f_{52}(6)$ | → | 0 | 0 | 0 | 0 | 0 |

[Table-2: Shows the process of exhaustive testing to draw the STD for Rule 52 UCA4NB]

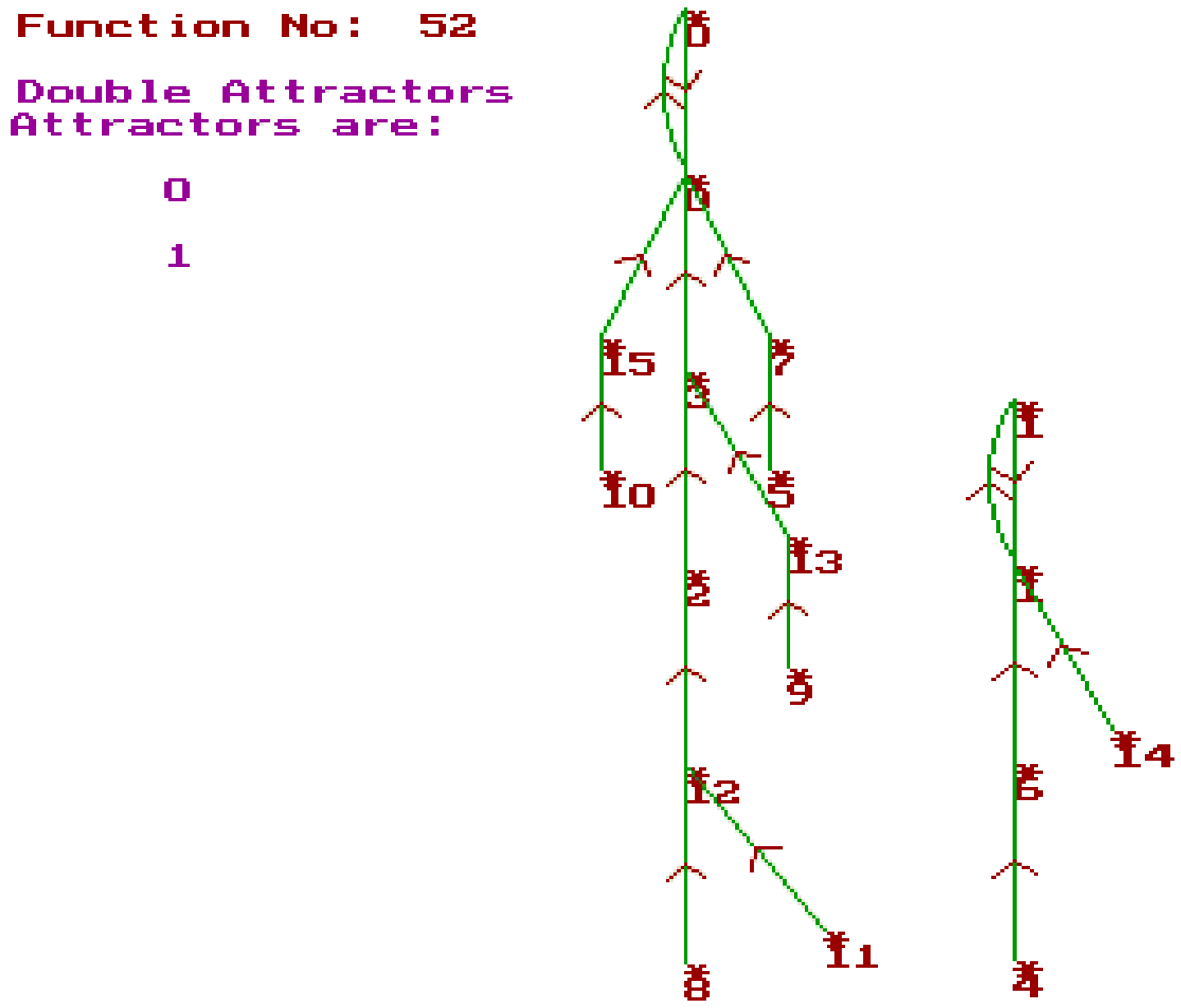

[Fig-1: The complete STD for Rule 52 UCA4NB]

As STD of a CA rule is necessarily a directed graph so the obvious way to represent an STD is to use an adjacency matrix. But, an efficient way to represent an STD is to use a one-dimensional array as the data structure. This requires only storing the successor information for all possible states. Fig-2 shows the successor array for the STD of Rule 52 UCA4NB.

| | | | | | | | | | | | | | | | | |
|---|---|---|---|---|---|---|---|---|---|---|---|---|---|---|---|---|
| Successors | 0 | 1 | 3 | 0 | 6 | 7 | 1 | 0 | 12 | 13 | 15 | 12 | 2 | 3 | 1 | 0 |

[Fig-2: Array representation of an STD for Rule 52 UCA4NB]

In a similar way the STD for Rule 52 UCA4PB can be computed and stored in a successor array and this is shown in table 3 followed by fig-3.

| States | $x_1^t$ | $x_2^t$ | $x_3^t$ | $x_4^t$ | $f_{52}$ | $x_1^{t+1}$ | $x_2^{t+1}$ | $x_3^{t+1}$ | $x_4^{t+1}$ | ≡ | $x_1^{t+1}$ | $x_2^{t+1}$ | $x_3^{t+1}$ | $x_4^{t+1}$ | Successors |
|---|---|---|---|---|---|---|---|---|---|---|---|---|---|---|---|
| 0 | 0 | 0 | 0 | 0 | → | $f_{52}(0)$ | $f_{52}(0)$ | $f_{52}(0)$ | $f_{52}(0)$ | → | 0 | 0 | 0 | 0 | 0 |
| 1 | 0 | 0 | 0 | 1 | → | $f_{52}(4)$ | $f_{52}(0)$ | $f_{52}(1)$ | $f_{52}(2)$ | → | 1 | 0 | 0 | 1 | 9 |
| 2 | 0 | 0 | 1 | 0 | → | $f_{52}(0)$ | $f_{52}(1)$ | $f_{52}(2)$ | $f_{52}(4)$ | → | 0 | 0 | 1 | 1 | 3 |
| 3 | 0 | 0 | 1 | 1 | → | $f_{52}(4)$ | $f_{52}(1)$ | $f_{52}(3)$ | $f_{52}(6)$ | → | 1 | 0 | 0 | 0 | 8 |
| 4 | 0 | 1 | 0 | 0 | → | $f_{52}(1)$ | $f_{52}(2)$ | $f_{52}(4)$ | $f_{52}(0)$ | → | 0 | 1 | 1 | 0 | 6 |
| 5 | 0 | 1 | 0 | 1 | → | $f_{52}(5)$ | $f_{52}(2)$ | $f_{52}(5)$ | $f_{52}(2)$ | → | 1 | 1 | 1 | 1 | 15 |
| 6 | 0 | 1 | 1 | 0 | → | $f_{52}(1)$ | $f_{52}(3)$ | $f_{52}(6)$ | $f_{52}(4)$ | → | 0 | 0 | 0 | 1 | 1 |
| 7 | 0 | 1 | 1 | 1 | → | $f_{52}(5)$ | $f_{52}(3)$ | $f_{52}(7)$ | $f_{52}(6)$ | → | 1 | 0 | 0 | 0 | 8 |
| 8 | 1 | 0 | 0 | 0 | → | $f_{52}(2)$ | $f_{52}(4)$ | $f_{52}(0)$ | $f_{52}(1)$ | → | 1 | 1 | 0 | 0 | 12 |
| 9 | 1 | 0 | 0 | 1 | → | $f_{52}(6)$ | $f_{52}(4)$ | $f_{52}(1)$ | $f_{52}(3)$ | → | 0 | 1 | 0 | 0 | 4 |
| 10 | 1 | 0 | 1 | 0 | → | $f_{52}(2)$ | $f_{52}(5)$ | $f_{52}(2)$ | $f_{52}(5)$ | → | 1 | 1 | 1 | 1 | 15 |
| 11 | 1 | 0 | 1 | 1 | → | $f_{52}(6)$ | $f_{52}(5)$ | $f_{52}(3)$ | $f_{52}(7)$ | → | 0 | 1 | 0 | 0 | 4 |
| 12 | 1 | 1 | 0 | 0 | → | $f_{52}(3)$ | $f_{52}(6)$ | $f_{52}(4)$ | $f_{52}(1)$ | → | 0 | 0 | 1 | 0 | 2 |
| 13 | 1 | 1 | 0 | 1 | → | $f_{52}(7)$ | $f_{52}(6)$ | $f_{52}(5)$ | $f_{52}(3)$ | → | 0 | 0 | 1 | 0 | 2 |
| 14 | 1 | 1 | 1 | 0 | → | $f_{52}(3)$ | $f_{52}(7)$ | $f_{52}(6)$ | $f_{52}(5)$ | → | 0 | 0 | 0 | 1 | 1 |
| 15 | 1 | 1 | 1 | 1 | → | $f_{52}(7)$ | $f_{52}(7)$ | $f_{52}(7)$ | $f_{52}(7)$ | → | 0 | 0 | 0 | 0 | 0 |

[Table-3: Shows the process to draw the STD for Rule 52 UCA4PB]

So in 4-bit the complete STD and the successor array for rule 52 periodic boundary is as follows:

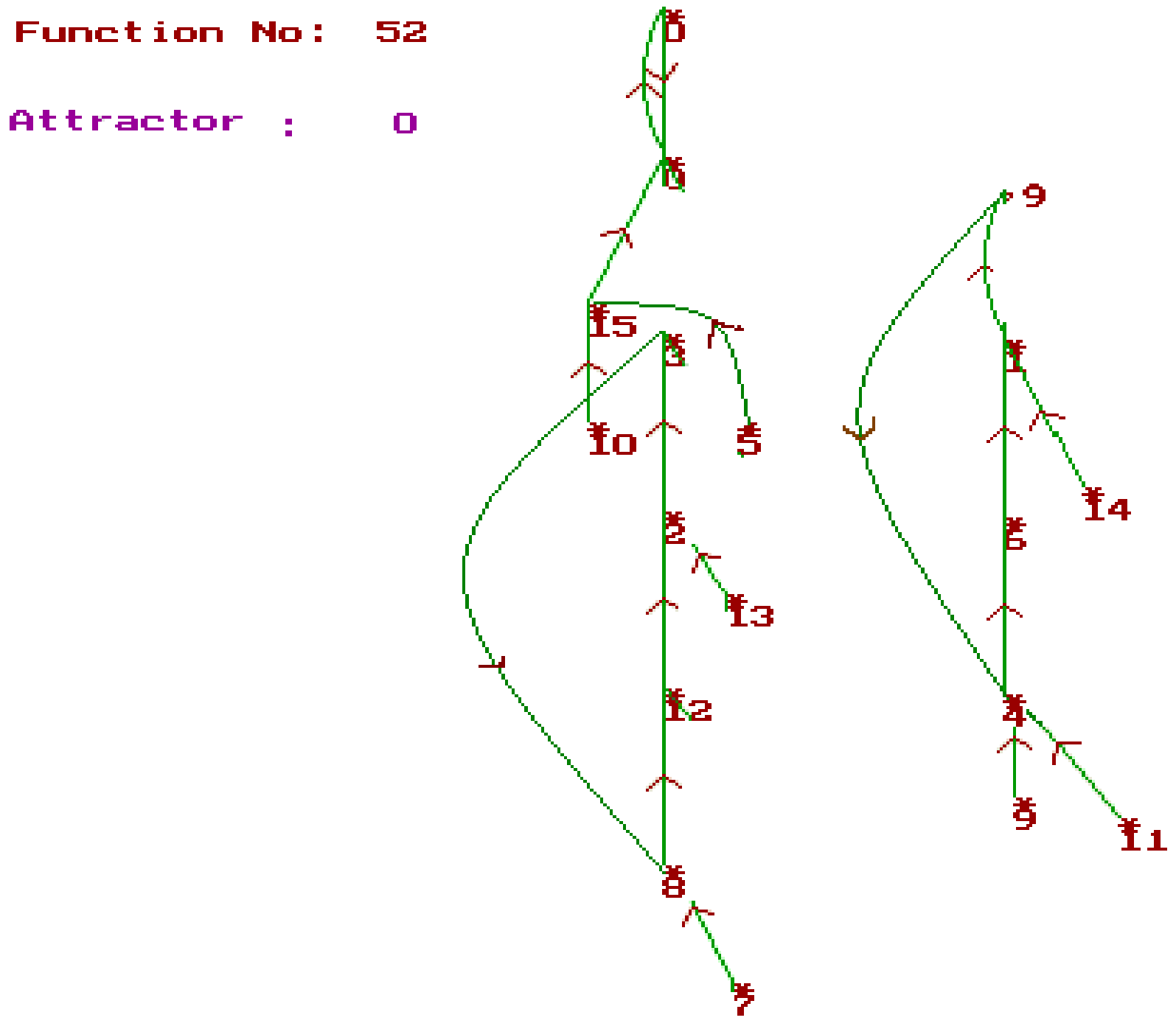

| | | | | | | | | | | | | | | | | |
|---|---|---|---|---|---|---|---|---|---|---|---|---|---|---|---|---|
| Successors | 0 | 9 | 3 | 8 | 6 | 15 | 1 | 8 | 12 | 4 | 15 | 4 | 2 | 2 | 1 | 0 |

[Fig-3: The complete STD and the successor array for Rule 52, UCA4PB using algorithm 3.1]

It is to be noted from fig 1 and fig 3 that due to the changing boundary condition, the dynamics of NB getting changed on the states 1, 3, 5, 7, 9, 11 and 13 for PB. In other words, only the edges concerning those states are getting changed. Also it affects the number of attractors for the same rule. For rule 52 NB it is 2 where as incase of PB it is 1. The complete study of this kind of dynamics has been initiated in our paper [18] and can also be taken up in future efforts.

The STD of n bit CA hybrid rules can also be drawn in a similar way. Only change is to apply different rules at different cells to get the next state. For k-variable Boolean functions (equivalently higher dimensional CA or 1D CA with k- neighborhood), the rule length is $2^k$ in the truth table and similar procedure will work to draw its STD.

To draw the STD of an arbitrary one-dimensional rule (uniform or hybrid) for n-bit CA using the above process requires ($4n.2^n$) READ and WRITE operations. This is because each cell requires 3 READ operations to collect its neighbor's information then 1 WRITE operation to update its own state as shown in fig 4 and there are $2^n$ states having n cells each. This assumes all these READ or WRITE operations take equal amount of time.

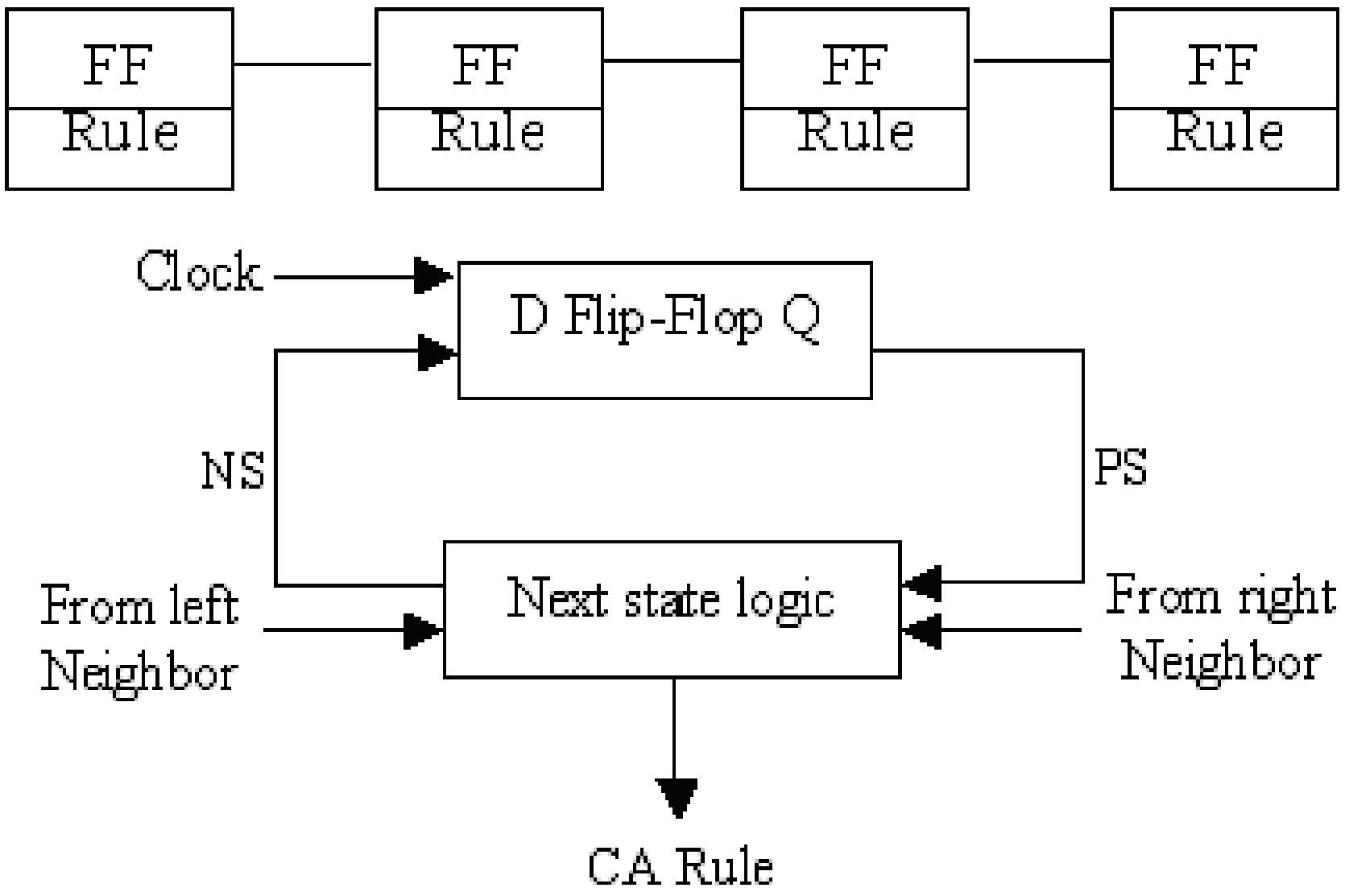

[Fig-4: Shows a 4-cell CA and the way each cell change its state using 3 neighborhoods CA rule]

It should be made clear that the time complexity defined here is obviously a software issue and should not be confused with the time complexity of a dedicated Cellular Automata Machine (CAM) which is basically a hardware device. In the CAM a single clock cycle is required for each state (i.e. an n-bit CA) to get its successor. Thus the time required for all the $2^n$ states to draw the complete STD is also $2^n$ and the space requirement in this case is (8+n); where 8 is the number of bits required to store the 1D CA rule and n is the length of CA.

Both these issues (hardware complexity and software complexity) are highly essential because a dedicated CAM architecture is not commonly available in the market and also very costly. So for the purpose of CA analysis researchers are drawing the STD by running a program using a standard PC or they are using certain simulation software's like DDLab [17] where the readymade codes for these algorithms are available.

Next an algorithm is developed to draw the complete STD for an arbitrary CA rule by which one can avoid all the READ operations involved at each cell in the exhaustive testing procedure. A new CAM architecture is proposed which supports this algorithm. Several issues regarding cost of the architecture and the novelty of this algorithm are also discussed.

### 3.1 Algorithms to draw the STD of null and periodic boundary CA with a given rule X and length of the CA n

Observing the bit patterns in table 2, table 3 and table 4, we develop the following algorithm for a given CA rule $x$ and length of the CA n. While drawing the STD in null and periodic boundary CA, they differ only in the boundary column bits in their successor states.

**Algorithm 3.1**

**Input:** Given a rule number $x$ and length of the CA n.
**Step-1:** Construct a matrix of dimension $(2^n \times n)$. Where $i^{th}$ row in this matrix correspond to, the input state $i$ in decimal form where $0 \le i \le 2^n - 1$.

**Step-2: [Boundary column]**

***if Null boundary CA then***
**(a)** The right most bit column or $x_n$ can be filled from the truth table of rule $x$ ( $f_x(0)$, $f_x(2)$, $f_x(4)$, $f_x(6)$ ). This 4-bit string is repeated for the entire column as shown in table 2 for rule 52.

**(b)** The left most bit column or $x_1$ can be filled up using ( $f_x(0)$, $f_x(1)$, $f_x(2)$, $f_x(3)$ ). And each $f_x(i)$ for $i$ =0, 1, 2, 3 is repeated for $2^n/4 = 2^{n-2}$ times as shown in table 2.

***else if Periodic boundary CA then***
**(a)** The right most bit column or $x_n$ can be filled from the truth table of rule $x$ ( $f_x(0)$, $f_x(2)$, $f_x(4)$, $f_x(6)$ ). This sequence is repeated for $2^n/2 = 2^{n-1}$ bits. Rest $2^{n-1}$ bits are a repetition of the 4-bit string ( $f_x(1)$, $f_x(3)$, $f_x(5)$, $f_x(7)$ ) as shown in table 3.

**(b)** The left-most column or $x_1$ can be filled up from the output of the given rule for the input string sequence corresponding to (( $f_x(0)$, $f_x(4)$ ), ( $f_x(1)$, $f_x(5)$ ), ( $f_x(2)$, $f_x(6)$ ), ( $f_x(3)$, $f_x(7)$ )) and each group is repeated for $2^n/4 = 2^{n-2}$ times as shown in table 3.

**Step-3: [Other column]:** See table 2 and table 3 for its correctness.
$x_{n-1}$ column can be filled up by putting the rule number in binary form for $2^n/8$ times successively.
$x_{n-2}$ column can be filled up by repeating each bit of the rule number 2 times successively.
$x_{n-3}$ column can be filled up by repeating each bit of the rule number 4 times successively.
.
.
.
$x_2$ column can be filled up, by repeating each bit of the rule number successively for $2^n/8$ times.

**Step-4:** find the decimal values of each n-bit string in the output matrix and store the result in an array of size $2^n$.

Drawing STD for any CA rule-using algorithm 3.1 does not involve any READ operation and requires $n2^n$ operations. The correctness of this algorithm is as follows.

**Correctness of algorithm 3.1:**

Step 1 of algorithm 3.1 is the construction of a table to store all the $2^n$ possible reachable states in the STD considering the CA length n. Step-2 and Step 3 contains the functional values in different cell positions after the CA rule $f_x$ is applied to all possible n-bit states arranged in ascending order of their decimal values. It is a well-known fact that if all possible n-bit states from $00...0$, $\ldots 11...1$ are arranged in ascending order in the table constructed in Step 1 then the right most column (or MSB or $x_n$) bits are in a sequence like $0101...01$ (alternate 0's and 1's of length $2^n$). Similarly the $2^n$ number of bits in the (n-1)$^{th}$ column corresponding to $x_{n-1}$ are in a sequence like $0011...11$ (two 0's followed by two 1's…) etc. and the bits in the 1$^{st}$ column or $x_1$ $^{th}$ column are in a sequence like $00...01...11$ (containing $2^{n-1}$ number of 0's followed by $2^{n-1}$ number of 1's). Further, in case of null boundary CA the left neighbor of $x_1$ and the right neighbor of $x_n$ are taken as 0. Thus padding two extra columns in the left and right sides of the table as all 0's and applying the 1D CA rule $f_x$ locally to each cell one can easily verify Steps 2 and 3 to get the functional values $f_x(p_i)\ for\ 0 \le p_i \le 7$ and $i = 1,2...n$ at different cell positions. As the bit patterns in $x_1$ and $x_n$ $^{th}$ columns are fixed after sorting the states, thus in periodic boundary CA one can easily get a different set of functional values only in the boundary columns as shown in Step 2. This is because NB and PB they differ only in their boundary cells. This proves the correctness of the algorithm for arbitrary n.

**CAM that supports algorithm 3.1:**

A CAM architecture, which supports algorithm 3.1, is shown in fig 5 and it works as follows.

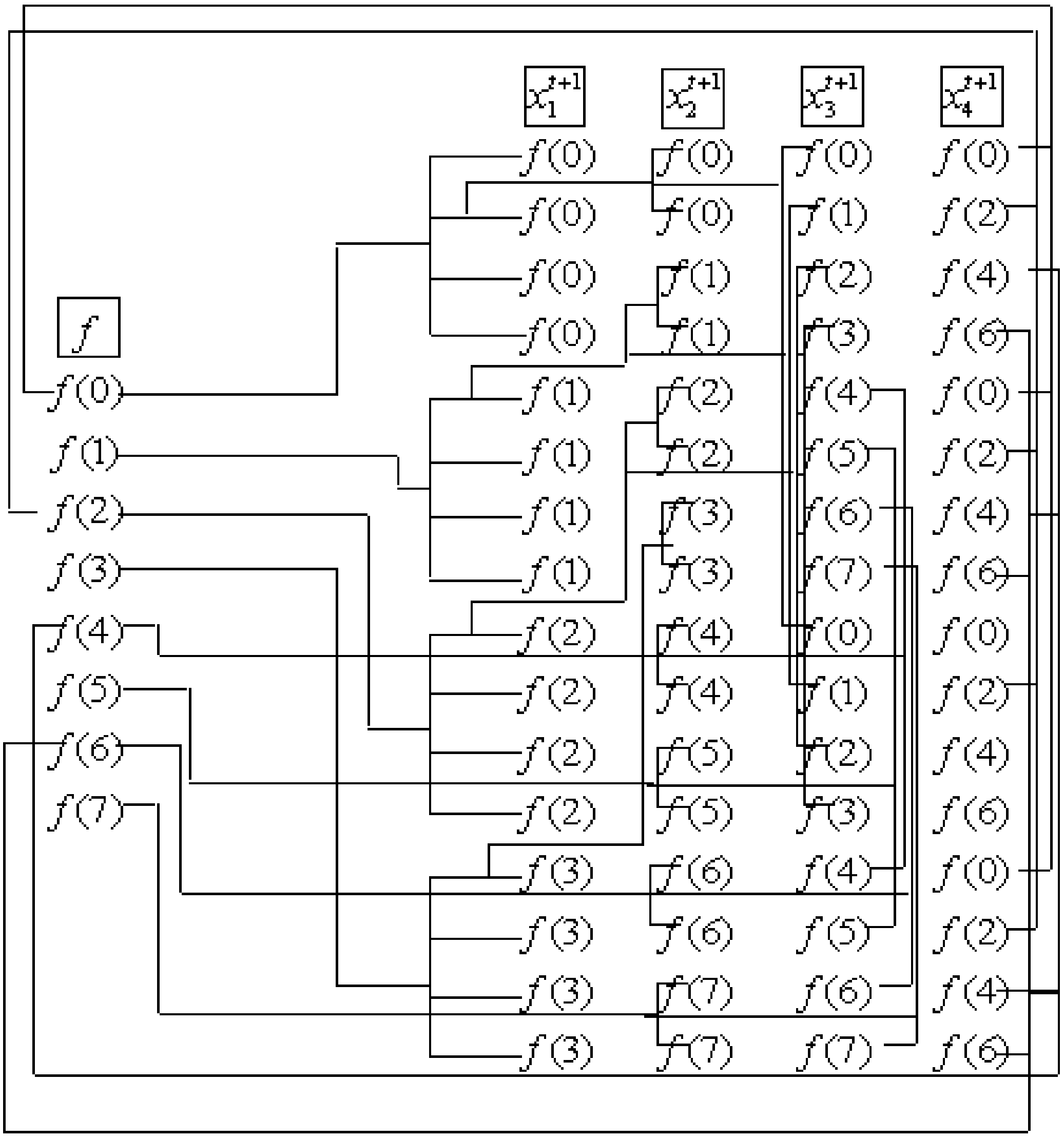

[Fig-5: CAM for an arbitrary UCA Rule X, in Null boundary using 4-bit CA for algorithm 3.1]

Consider an example of a 1-D 3-neighborhood UCA. Here 256 UCA rules are possible. As shown in fig-5 at first an arbitrary CA rule $f$ is loaded as an 8-bit string. Because all the CA cells are connected through wires, the 8 functional values $f(0), f(1),...f(7)$ are immediately transferred to their corresponding cell positions by a single load instruction given at $f$ and produce all the successors in the state space in ascending order. Thus CAM can draw the complete STD in a single clock cycle. This method does not involve any READ operation associated with the conventional CA procedure as shown in fig-4 (cells change their own states by reading the states of their neighborhoods). Here only loading a CA rules produces its STD. One drawback of this CAM is the memory requirement to store all the $2^n$ states containing n cells each and a huge number of connections (or wires) required between the functional values and the corresponding CA cells. For 256 UCA the space requirement for this CAM is ($8+n2^n$); where 8 is the number of bits required to store the 1D CA rule and n is the length of CA.

As in case of HCA different CA rules are applied to different cell positions, thus for n-bit CA in the worst case n dedicated functional units say $f_1, f_{2,} ..., f_n$ can be used containing 8 bits each. The corresponding connections through wires can be made between different functional units to different column positions such that $i^{th}$ functional unit $f_i$ is connected to the successor column $x_i^{t+1}$ for different values of $i = 1, 2, ..., n$. Here also after loading all the functions in different functional units, only 1 clock cycle is used to get all the successors in the state space. But, the space requirement in this case is ($8n+n2^n$) and the wiring costs are also more than the CAM used for UCA. The CAM for UCA and/or HCA with more than 3 neighborhoods can also be designed in a similar way.

**Characteristics of STD obtained from algorithm 3.1**

For any rule one can find out the output array of length $2^n$ by the above algorithm and analysis of this output array gives several useful results.

1. **STD:** One can draw the complete STD by drawing directed arrows from i to A[i]. Here we do not need any input string and also we do not need to apply the particular rule to each bit. Only knowing the rule number one can draw its STD.

2. **Reachable and non-reachable states:** the numbers those are not present in the resultant array are the non-reachable states. For example from the output array of fig 2 gives 7 non-reachable states and those are 4, 14, 10, 8, 11, 9 and 5, which can be verified for the STD shown in fig 1.

3. **Predecessors:** If a state appears k-number of times in the resultant array then it signifies that it has got k-number of predecessors for some positive integer k. State 0 has 4-predecessors, state 1 has 3-predecessors, state 3 has 2-predecessors and so on in fig 2. Here predecessor we mean the immediate predecessor.

4. **Self loop:** If $i = A[i]$ then state $i$ forms a self loop. From fig 2 states 0 and 1 contain self-loops.

5. **Point attractor:** A self-loop having k-predecessors for some positive integer k, is a point attractor. In fig-2 states 0 and 1 are point attractors.

**Remark:** This algorithm is more general for arbitrary n-bit CA and it is valid for both uniform/hybrid, null/periodic boundary CA rules. The result is also consistent for 1D CA of any arbitrary neighborhood $k$ .

Our next discussion further reduces the time and space complexity of the above algorithm. Observing the functional values row wise in table 2 and table 3 in null and periodic boundary conditions a recursive algorithm can be designed as follows.

**Algorithm 3.2**

**Input:** Given a rule number X and length of the CA n.

**Step-1:**
We are constructing trees within this step by the recursive procedures for null boundary as well as periodic boundary as follows. Here any reachable state in the STD is an n-bit string of the form $x = x_1x_2x_3...x_n$ where each bit $x_i$ is a node of the tree and can be computed from $x_{i-1}$. Here $x_1$ is the root and $0 \le p_i \le 7 \quad for\, i = 1,2,...,n$ is a position in decimal in the truth table of 3-variable Boolean function $f$ . Basically $p_i$ is a decimal value of all possible 3-bit string from 000, 001… and 111 in the truth table in case of 1D CA.

### *for Null boundary CA*

Step 1: Draw four root nodes, labeled 0, 1, 2, and 3.
Step $i$ ( $i = 2,3,...,n-1$): for each node drawn in step $(i-1)$ labeled $p_{i-1}$, draw two children labeled $(2p_{i-1}) \bmod 8$ and $(((2p_{i-1}) \bmod 8) + 1)$ respectively.
Step $n$: For each node drawn in Step $(n-1)$ labeled $p_{n-1}$, draw a child labeled $(2p_{n-1}) \bmod 8$ .

### *for Periodic boundary CA*

Step1: Draw four root nodes, labeled (0, 4), (1, 5), (2, 6), and (3, 7).
Step $i$ ( $i = 2,3,...,n-1$): for each node drawn in step $(i-1)$ labeled $p_{i-1}$, draw two children labeled $(2p_{i-1}) \bmod 8$ and $(((2p_{i-1}) \bmod 8) + 1)$ respectively.
Step $n$: For each node drawn in Step $(n-1)$ labeled $p_{n-1}$, draw a child labeled $((2p_{n-1}) \bmod 8 + x_1)$ .

**Remark:** In case of null boundary a single functional value is stored at the root node where as in periodic boundary two functional values are stored at the root as shown in fig 6 for n=4.

**Step-2:**
In null boundary, The DFS traversal of this binary tree and on keeping the label information from the root node to leaf nodes produces all possible successors in ascending order corresponding to the CA states starting from 0 to $2^{n-1}$. But in case of periodic boundary same DFS traversal is used but at the root node the functional information is collected alternately.

For example, using step 2 of null boundary CA as shown in fig 6 (a) for n=4 one can obtain the string ( $f(0)$ , $f(0)$ , $f(0)$ , $f(0)$ ) as the successor of state 0=0000, ( $f(0)$ , $f(0)$ , $f(1)$ , $f(2)$ ) is the successor of state 1=0001, ( $f(0)$ , $f(1)$ , $f(2)$ , $f(4)$ ) is the successor of state 2=0010 and so on. In periodic boundary, ( $f(0)$ , $f(0)$ , $f(0)$ , $f(0)$ ) is the successor of the state 0=0000, ( $f(4)$ , $f(0)$ , $f(1)$ , $f(2)$ ) is the successor of state 1=0001, ( $f(0)$ , $f(1)$ , $f(2)$ , $f(4)$ ) is the successor of state 2=0010, ( $f(4)$ , $f(1)$ , $f(3)$ , $f(6)$ ) is the successor of state 3=0011, ( $f(1)$ , $f(2)$ , $f(4)$ , $f(0)$ ,) is the successor of state 4=0100 and so on as shown in fig 6 (b).

**Correctness of algorithm 3.2:**
In the correctness of this algorithm 3.2 we can similarly follow the arguments stated in the correctness of the algorithm 3.1. The only difference lies in narrating the row structures instead of column structures. In algorithm 3.1 we utilize the fixed bit patterns of the different columns for arbitrary n, whereas in algorithm 3.2 we have obtained the recursive formulae (in step 1) by looking the patterns in row wise.

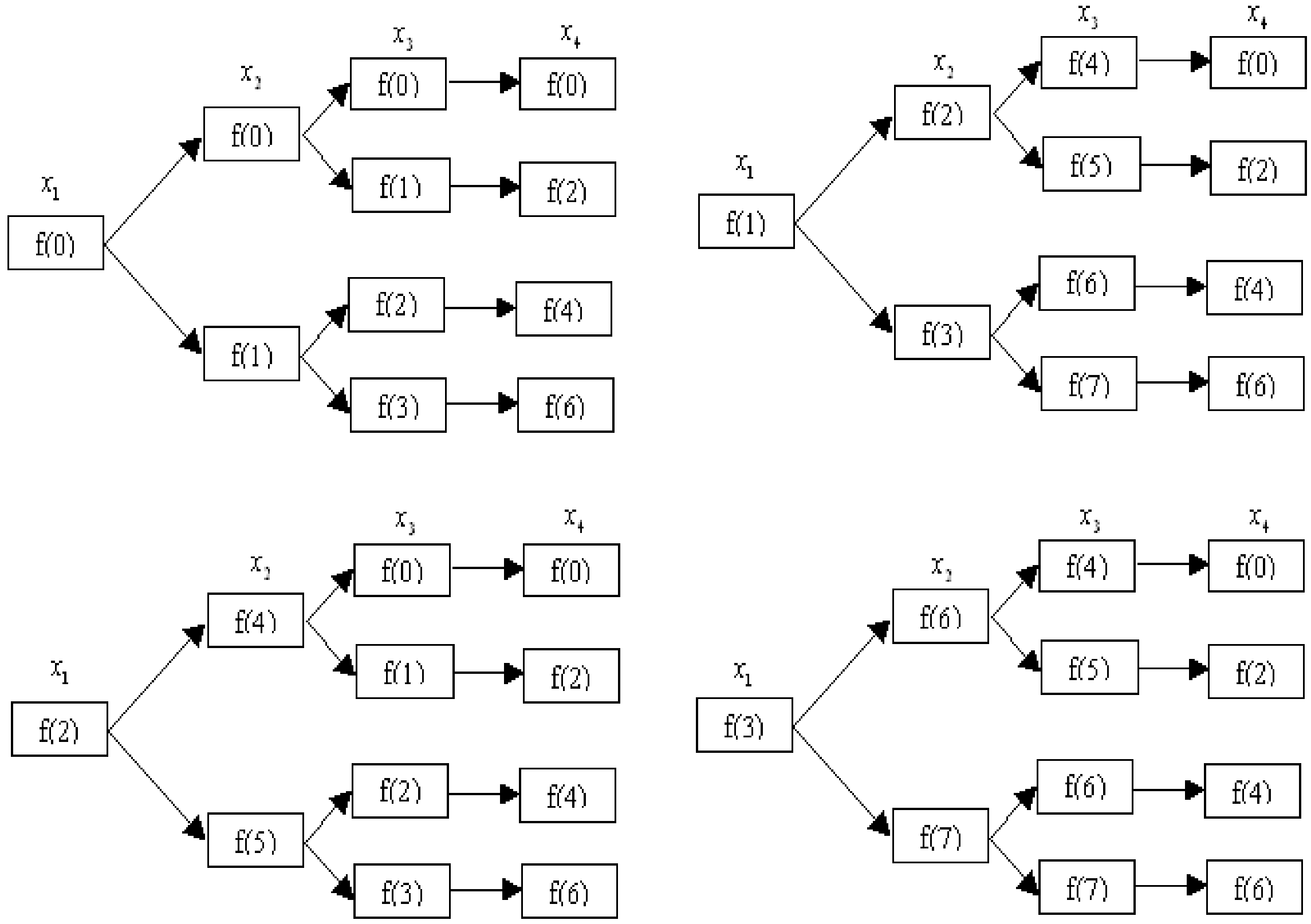

(a)

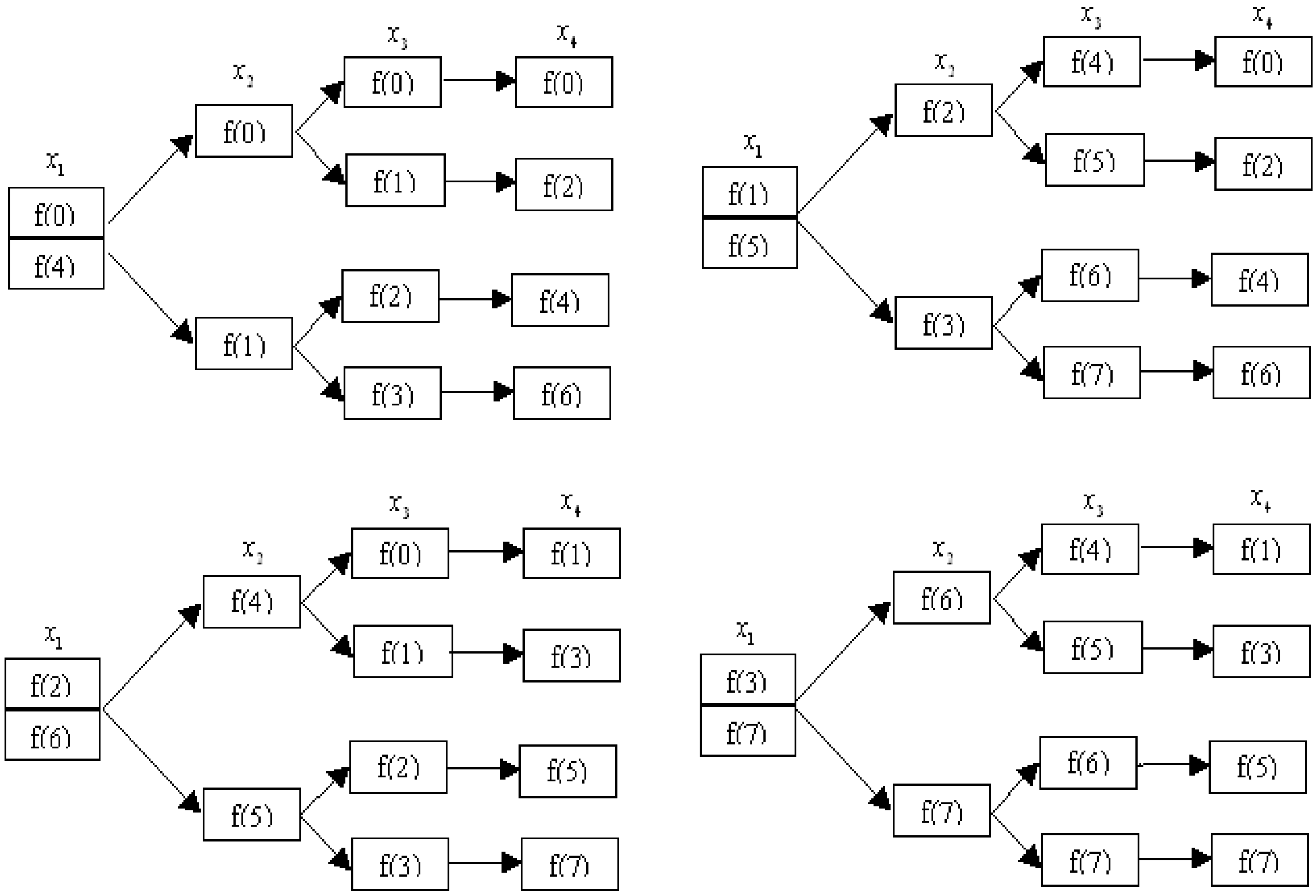

(b)

[Fig-6: (a) shows the binary tree representation for an arbitrary 1-D Rule X, and 4-bit CA in Null boundary condition and (b) shows the same for periodic boundary obtained using algorithm 3.2]

**CAM that supports algorithm 3.2:**

The CAM, which supports the recursive nature of algorithm 3.2 for null boundary UCA, is shown in fig 7. Here the modification is done at each column in the successor part by taking a single memory unit for same functional values and thus some memory units can be reduced. For example in case of 4-bit UCA CA as shown in fig 5, the four $f(0)$ values present in the first column $x_1^{t+1}$ can be combined to a single memory unit $f(0)$ and a single memory units are required for $f(1)$, $f(2)$ and $f(3)$ respectively. Similarly, memory units for other column positions in the successor part can also be set in the CAM, which is shown in fig 7. The space requirement of this CAM is also less than the previous CAM architecture and exactly it is equal to:

$$8+(2^2+2^3+...2^n)+2^n = 8+2^2(1+2+...2^{n-2})+2^n = 8+4(2^{n-1}-1)+2^n = 4+2^n+2^{n+1} = 4+3.2^n$$

This machine also works like the previous CAM architecture shown in fig 5 but the successors information are collected using the traversal procedure given in step 2 of algorithm 3.2. Because of the less memory unit, this architecture requires less number of connections and hence it is less costly with respect to wiring costs than the previous CAM architecture proposed in fig 5. Also in the hardware point of view this also takes a single clock cycle to get all the functional values required for the table to draw the complete STD. Thus in an overall sense this architecture is a better one than the previous architectures. The CAM for HCA can also be designed in a similar fashion by taking n functional units $f_1, f_{2,} ..., f_n$ as discussed earlier and in this case the space requirement is (8n+3.2$^n$-4). The CAM for UCA and/or HCA with more than 3 neighborhoods can also be designed in a similar way.

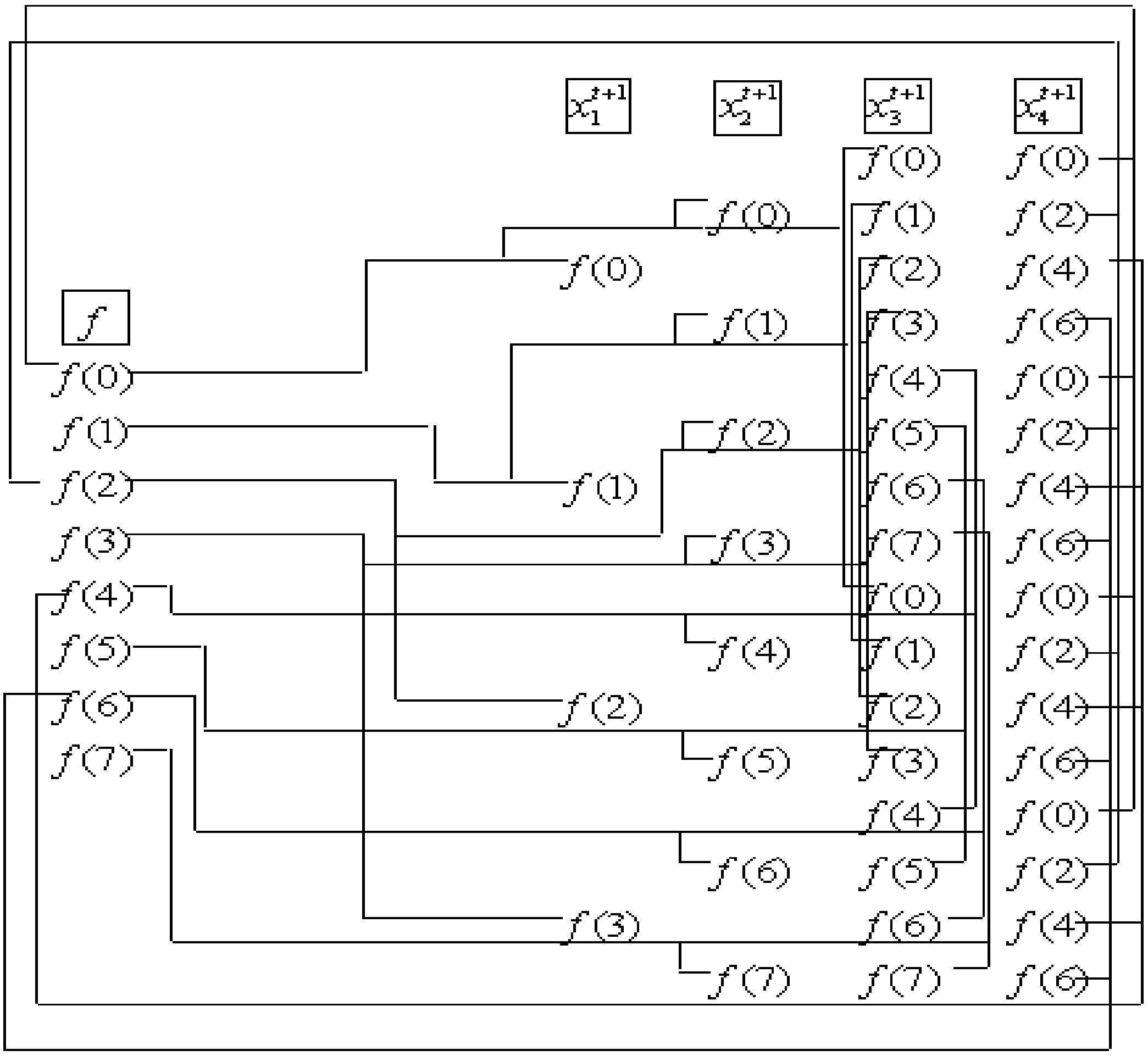

[Fig-7: CAM for arbitrary UCA Rule X, and 4-bit CA in Null boundary for algorithm 3.2]

This is to be noted that from algorithm 3.2, for finding the successor of any arbitrary state we need not compute the whole STD. Instead we can easily find the successor of any arbitrary state using the following procedure.

**Procedure for finding Successor:** Given an arbitrary state $y = y_1 y_2 y_3 ... y_n$ in decimal and the CA rule $f$ finding its successor $x = x_1 x_2 x_3 ... x_n$ is easy using algorithm 3.2. First, one has to reach at the root node of the binary tree i.e. $x_1 = f(p_1)$ by computing $p_1 = 0 y_1 y_2$ in case of null boundary or $p_1 = y_{n-1} y_1 y_2$ in case of periodic boundary CA. Then other bits $x_2 = f(p_2), x_3 = f(p_3) ... x_n = f(p_n)$ can be computed using the recursive algorithm given in step 1.

Again using repeatedly the above successor finding procedure (Start from a CA state, find its successor, and find each successor's successor, and so on) we are landing up with one trajectory or a directed path, which does not necessarily exhaust the state-space. For example in fig 1, if we start in state 11 then we have the path $11 \rightarrow 12 \rightarrow 2 \rightarrow 3 \rightarrow 0 \leftrightarrow 0$. Obviously it does not cover the other branches of the connected components as can be obtained from direct reverse algorithm.

## 4. Conclusion and future work

In this paper we have addressed several issues on drawing the STD's for arbitrary CA rules (uniform and/or hybrid) of length n and also contribute some new algorithms to the existing literature. One of the requirements to draw the STD for an arbitrary CA rule is to predict some of the characteristics (reachable/non-reachable states, loops, attractors etc.) of an n-length CA for sufficiently large values of n by observing the STD of CA rule for small values of n. Thus to understand the global dynamics of non-linear CA the newly introduced algorithms and their CAM implementations are obviously helpful.

## Appendix 1

Here a table is constructed on using algorithm 3.1 for Rule 52 UCA5NB condition. The actual table size is $2^5 \times 5 = 32 \times 5 = 160$. Hence 160 memory units are required to store all the functional values $f_{52}(p_i)$ for $0 \le p_i \le 7 \quad for\, i = 1, 2, ..., n$. A portion of the binary tree corresponding to this table is shown in fig 8 and the complete STD is shown in fig 9.

| $x_1^{t+1}$ | $x_2^{t+1}$ | $x_3^{t+1}$ | $x_4^{t+1}$ | $x_5^{t+1}$ | $\equiv$ | $x_1^{t+1}$ | $x_2^{t+1}$ | $x_3^{t+1}$ | $x_4^{t+1}$ | $x_5^{t+1}$ |
|---|---|---|---|---|---|---|---|---|---|---|
| $f_{52}(0)$ | $f_{52}(0)$ | $f_{52}(0)$ | $f_{52}(0)$ | $f_{52}(0)$ | $\rightarrow$ | 0 | 0 | 0 | 0 | 0 |
| $f_{52}(0)$ | $f_{52}(0)$ | $f_{52}(0)$ | $f_{52}(1)$ | $f_{52}(2)$ | $\rightarrow$ | 0 | 0 | 0 | 0 | 1 |
| $f_{52}(0)$ | $f_{52}(0)$ | $f_{52}(1)$ | $f_{52}(2)$ | $f_{52}(4)$ | $\rightarrow$ | 0 | 0 | 0 | 1 | 1 |
| $f_{52}(0)$ | $f_{52}(0)$ | $f_{52}(1)$ | $f_{52}(3)$ | $f_{52}(6)$ | $\rightarrow$ | 0 | 0 | 0 | 0 | 0 |
| $f_{52}(0)$ | $f_{52}(1)$ | $f_{52}(2)$ | $f_{52}(4)$ | $f_{52}(0)$ | $\rightarrow$ | 0 | 0 | 1 | 1 | 0 |
| $f_{52}(0)$ | $f_{52}(1)$ | $f_{52}(2)$ | $f_{52}(5)$ | $f_{52}(2)$ | $\rightarrow$ | 0 | 0 | 1 | 1 | 1 |
| $f_{52}(0)$ | $f_{52}(1)$ | $f_{52}(3)$ | $f_{52}(6)$ | $f_{52}(4)$ | $\rightarrow$ | 0 | 0 | 0 | 0 | 1 |
| $f_{52}(0)$ | $f_{52}(1)$ | $f_{52}(3)$ | $f_{52}(7)$ | $f_{52}(6)$ | $\rightarrow$ | 0 | 0 | 0 | 0 | 0 |
| $f_{52}(1)$ | $f_{52}(2)$ | $f_{52}(4)$ | $f_{52}(0)$ | $f_{52}(0)$ | $\rightarrow$ | 0 | 1 | 1 | 0 | 0 |
| $f_{52}(1)$ | $f_{52}(2)$ | $f_{52}(4)$ | $f_{52}(1)$ | $f_{52}(2)$ | $\rightarrow$ | 0 | 1 | 1 | 0 | 1 |
| $f_{52}(1)$ | $f_{52}(2)$ | $f_{52}(5)$ | $f_{52}(2)$ | $f_{52}(4)$ | $\rightarrow$ | 0 | 1 | 1 | 1 | 1 |
| $f_{52}(1)$ | $f_{52}(2)$ | $f_{52}(5)$ | $f_{52}(3)$ | $f_{52}(6)$ | $\rightarrow$ | 0 | 1 | 1 | 0 | 0 |
| $f_{52}(1)$ | $f_{52}(3)$ | $f_{52}(6)$ | $f_{52}(4)$ | $f_{52}(0)$ | $\rightarrow$ | 0 | 0 | 0 | 1 | 0 |
| $f_{52}(1)$ | $f_{52}(3)$ | $f_{52}(6)$ | $f_{52}(5)$ | $f_{52}(2)$ | $\rightarrow$ | 0 | 0 | 0 | 1 | 1 |
| $f_{52}(1)$ | $f_{52}(3)$ | $f_{52}(7)$ | $f_{52}(6)$ | $f_{52}(4)$ | $\rightarrow$ | 0 | 0 | 0 | 0 | 1 |
| $f_{52}(1)$ | $f_{52}(3)$ | $f_{52}(7)$ | $f_{52}(7)$ | $f_{52}(6)$ | $\rightarrow$ | 0 | 0 | 0 | 0 | 0 |
| $f_{52}(2)$ | $f_{52}(4)$ | $f_{52}(0)$ | $f_{52}(0)$ | $f_{52}(0)$ | $\rightarrow$ | 1 | 1 | 0 | 0 | 0 |
| $f_{52}(2)$ | $f_{52}(4)$ | $f_{52}(0)$ | $f_{52}(1)$ | $f_{52}(2)$ | $\rightarrow$ | 1 | 1 | 0 | 0 | 1 |
| $f_{52}(2)$ | $f_{52}(4)$ | $f_{52}(1)$ | $f_{52}(2)$ | $f_{52}(4)$ | $\rightarrow$ | 1 | 1 | 0 | 1 | 1 |
| $f_{52}(2)$ | $f_{52}(4)$ | $f_{52}(1)$ | $f_{52}(3)$ | $f_{52}(6)$ | $\rightarrow$ | 1 | 1 | 0 | 0 | 0 |
| $f_{52}(2)$ | $f_{52}(5)$ | $f_{52}(2)$ | $f_{52}(4)$ | $f_{52}(0)$ | $\rightarrow$ | 1 | 1 | 1 | 1 | 0 |
| $f_{52}(2)$ | $f_{52}(5)$ | $f_{52}(2)$ | $f_{52}(5)$ | $f_{52}(2)$ | $\rightarrow$ | 1 | 1 | 1 | 1 | 1 |
| $f_{52}(2)$ | $f_{52}(5)$ | $f_{52}(3)$ | $f_{52}(6)$ | $f_{52}(4)$ | $\rightarrow$ | 1 | 1 | 0 | 0 | 1 |
| $f_{52}(2)$ | $f_{52}(5)$ | $f_{52}(3)$ | $f_{52}(7)$ | $f_{52}(6)$ | $\rightarrow$ | 1 | 1 | 0 | 0 | 0 |
| $f_{52}(3)$ | $f_{52}(6)$ | $f_{52}(4)$ | $f_{52}(0)$ | $f_{52}(0)$ | $\rightarrow$ | 0 | 0 | 1 | 0 | 0 |
| $f_{52}(3)$ | $f_{52}(6)$ | $f_{52}(4)$ | $f_{52}(1)$ | $f_{52}(2)$ | $\rightarrow$ | 0 | 0 | 1 | 0 | 1 |
| $f_{52}(3)$ | $f_{52}(6)$ | $f_{52}(5)$ | $f_{52}(2)$ | $f_{52}(4)$ | $\rightarrow$ | 0 | 0 | 1 | 1 | 1 |
| $f_{52}(3)$ | $f_{52}(6)$ | $f_{52}(5)$ | $f_{52}(3)$ | $f_{52}(6)$ | $\rightarrow$ | 0 | 0 | 1 | 0 | 0 |
| $f_{52}(3)$ | $f_{52}(7)$ | $f_{52}(6)$ | $f_{52}(4)$ | $f_{52}(0)$ | $\rightarrow$ | 0 | 0 | 0 | 1 | 0 |
| $f_{52}(3)$ | $f_{52}(7)$ | $f_{52}(6)$ | $f_{52}(5)$ | $f_{52}(2)$ | $\rightarrow$ | 0 | 0 | 0 | 1 | 1 |
| $f_{52}(3)$ | $f_{52}(7)$ | $f_{52}(7)$ | $f_{52}(6)$ | $f_{52}(4)$ | $\rightarrow$ | 0 | 0 | 0 | 0 | 1 |
| $f_{52}(3)$ | $f_{52}(7)$ | $f_{52}(7)$ | $f_{52}(7)$ | $f_{52}(6)$ | $\rightarrow$ | 0 | 0 | 0 | 0 | 0 |

[Table-4: Shows the functional values obtained using algorithm 3.1 to draw the complete STD for Rule 52 UCA5NB condition]

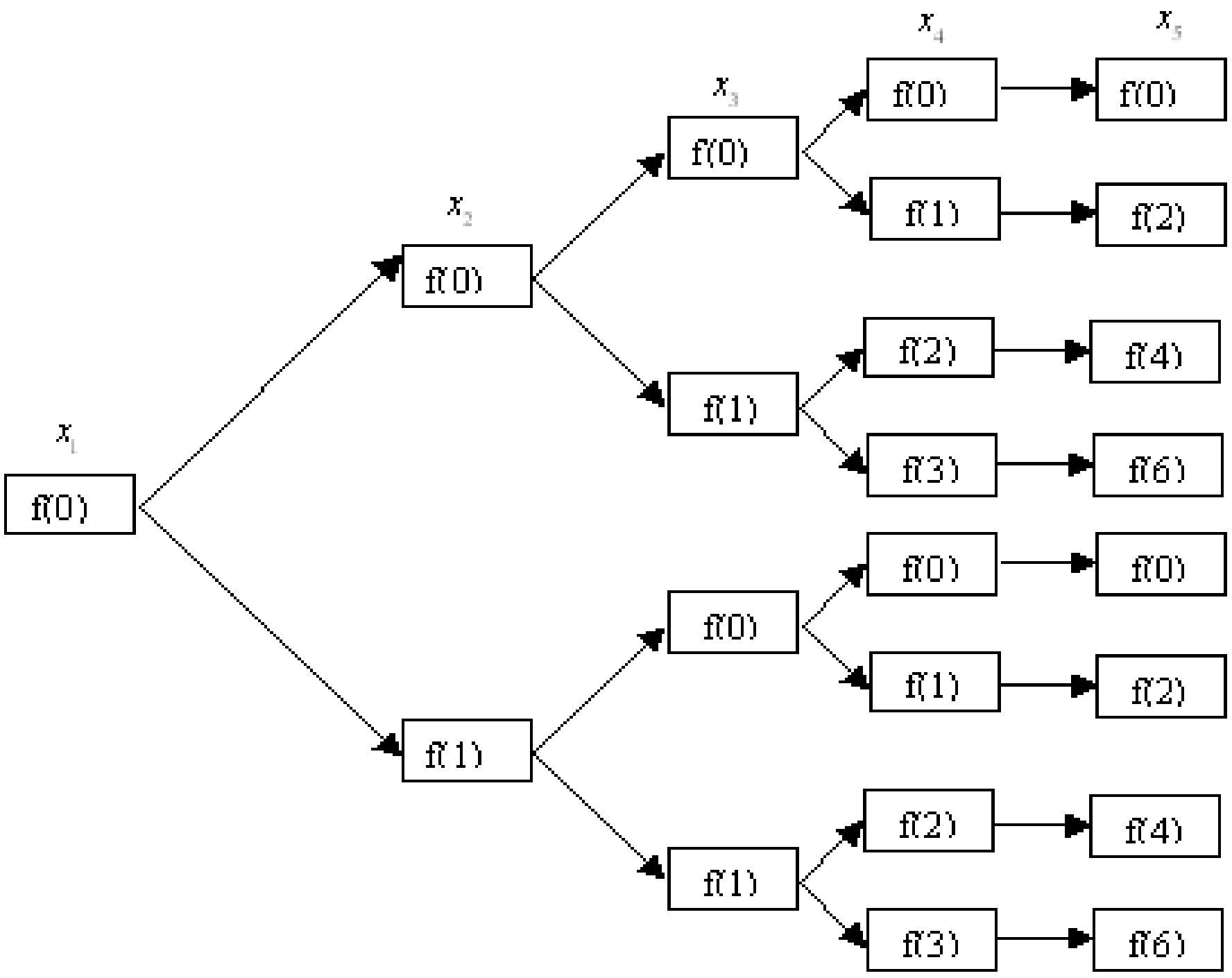

[Fig-8: Shows a portion of a binary tree representation for an arbitrary 1D uniform null boundary CA rule $f$ with length 5, obtained using algorithm 3.2. Here $x_i, i = 1, 2...5$ denotes the updated value of $x_i$ at time (t+1) after the CA rule $f$ is applied to it.]

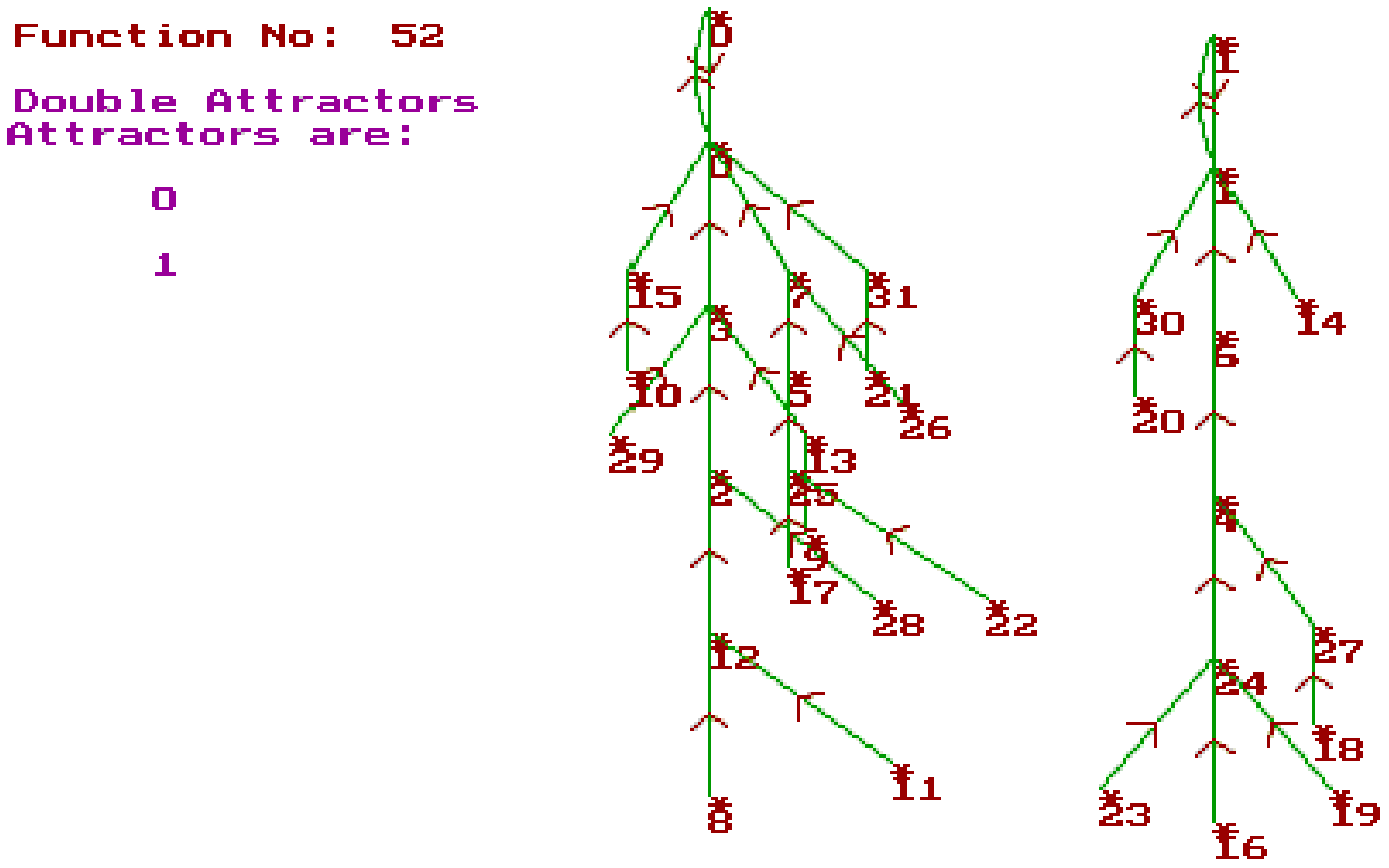

[Fig-9: The complete STD for Rule 52 UCA5NB]

## Appendix 2

Consider an example of 1D 3-neighborhood CA Rule 108. The truth table of this rule is shown in table 5. Two tables Table 6(a) and 6(b) are obtained on using algorithm 3.1 for Rule 108 both for UCA4NB and UCA4PB condition. Using algorithm 3.2 the binary trees corresponding to table 6(a) and 6(b) are shown in fig 10(a) and 10(b). Both the STDs for null and periodic boundary are shown in fig 11(a) and (b).

| Dec. Value | X | Y | Z | $f_{108}$ |
|---|---|---|---|---|
| 0 | 0 | 0 | 0 | 0 |
| 1 | 0 | 0 | 1 | 0 |
| 2 | 0 | 1 | 0 | 1 |
| 3 | 0 | 1 | 1 | 1 |
| 4 | 1 | 0 | 0 | 0 |
| 5 | 1 | 0 | 1 | 1 |
| 6 | 1 | 1 | 0 | 1 |
| 7 | 1 | 1 | 1 | 0 |

[Table-5: Truth table for rule 108 in 3 variable Boolean function]

| States | $x_1^t$ | $x_2^t$ | $x_3^t$ | $x_4^t$ | $f_{108}$ | $x_1^{t+1}$ | $x_2^{t+1}$ | $x_3^{t+1}$ | $x_4^{t+1}$ | $\equiv$ | $x_1^{t+1}$ | $x_2^{t+1}$ | $x_3^{t+1}$ | $x_4^{t+1}$ | Successors |
|---|---|---|---|---|---|---|---|---|---|---|---|---|---|---|---|
| 0 | 0 | 0 | 0 | 0 | $\rightarrow$ | $f_{108}(0)$ | $f_{108}(0)$ | $f_{108}(0)$ | $f_{108}(0)$ | $\rightarrow$ | 0 | 0 | 0 | 0 | 0 |
| 1 | 0 | 0 | 0 | 1 | $\rightarrow$ | $f_{108}(0)$ | $f_{108}(0)$ | $f_{108}(1)$ | $f_{108}(2)$ | $\rightarrow$ | 0 | 0 | 0 | 1 | 1 |
| 2 | 0 | 0 | 1 | 0 | $\rightarrow$ | $f_{108}(0)$ | $f_{108}(1)$ | $f_{108}(2)$ | $f_{108}(4)$ | $\rightarrow$ | 0 | 0 | 1 | 0 | 2 |
| 3 | 0 | 0 | 1 | 1 | $\rightarrow$ | $f_{108}(0)$ | $f_{108}(1)$ | $f_{108}(3)$ | $f_{108}(6)$ | $\rightarrow$ | 0 | 0 | 1 | 1 | 3 |
| 4 | 0 | 1 | 0 | 0 | $\rightarrow$ | $f_{108}(1)$ | $f_{108}(2)$ | $f_{108}(4)$ | $f_{108}(0)$ | $\rightarrow$ | 0 | 1 | 0 | 0 | 4 |
| 5 | 0 | 1 | 0 | 1 | $\rightarrow$ | $f_{108}(1)$ | $f_{108}(2)$ | $f_{108}(5)$ | $f_{108}(2)$ | $\rightarrow$ | 0 | 1 | 1 | 1 | 7 |
| 6 | 0 | 1 | 1 | 0 | $\rightarrow$ | $f_{108}(1)$ | $f_{108}(3)$ | $f_{108}(6)$ | $f_{108}(4)$ | $\rightarrow$ | 0 | 1 | 1 | 0 | 6 |
| 7 | 0 | 1 | 1 | 1 | $\rightarrow$ | $f_{108}(1)$ | $f_{108}(3)$ | $f_{108}(7)$ | $f_{108}(6)$ | $\rightarrow$ | 0 | 1 | 0 | 1 | 5 |
| 8 | 1 | 0 | 0 | 0 | $\rightarrow$ | $f_{108}(2)$ | $f_{108}(4)$ | $f_{108}(0)$ | $f_{108}(0)$ | $\rightarrow$ | 1 | 0 | 0 | 0 | 8 |
| 9 | 1 | 0 | 0 | 1 | $\rightarrow$ | $f_{108}(2)$ | $f_{108}(4)$ | $f_{108}(1)$ | $f_{108}(2)$ | $\rightarrow$ | 1 | 0 | 0 | 1 | 9 |
| 10 | 1 | 0 | 1 | 0 | $\rightarrow$ | $f_{108}(2)$ | $f_{108}(5)$ | $f_{108}(2)$ | $f_{108}(4)$ | $\rightarrow$ | 1 | 1 | 1 | 0 | 14 |
| 11 | 1 | 0 | 1 | 1 | $\rightarrow$ | $f_{108}(2)$ | $f_{108}(5)$ | $f_{108}(3)$ | $f_{108}(6)$ | $\rightarrow$ | 1 | 1 | 1 | 1 | 15 |
| 12 | 1 | 1 | 0 | 0 | $\rightarrow$ | $f_{108}(3)$ | $f_{108}(6)$ | $f_{52}(4)$ | $f_{108}(0)$ | $\rightarrow$ | 1 | 1 | 0 | 0 | 12 |
| 13 | 1 | 1 | 0 | 1 | $\rightarrow$ | $f_{108}(3)$ | $f_{108}(6)$ | $f_{108}(5)$ | $f_{108}(2)$ | $\rightarrow$ | 1 | 1 | 1 | 1 | 15 |
| 14 | 1 | 1 | 1 | 0 | $\rightarrow$ | $f_{108}(3)$ | $f_{108}(7)$ | $f_{108}(6)$ | $f_{108}(4)$ | $\rightarrow$ | 1 | 0 | 1 | 0 | 10 |
| 15 | 1 | 1 | 1 | 1 | $\rightarrow$ | $f_{108}(3)$ | $f_{108}(7)$ | $f_{108}(7)$ | $f_{108}(6)$ | $\rightarrow$ | 1 | 0 | 0 | 1 | 9 |

(a)

| States | $x_1^t$ | $x_2^t$ | $x_3^t$ | $x_4^t$ | $f_{108}$ | $x_1^{t+1}$ | $x_2^{t+1}$ | $x_3^{t+1}$ | $x_4^{t+1}$ | $\equiv$ | $x_1^{t+1}$ | $x_2^{t+1}$ | $x_3^{t+1}$ | $x_4^{t+1}$ | Successors |
|---|---|---|---|---|---|---|---|---|---|---|---|---|---|---|---|
| 0 | 0 | 0 | 0 | 0 | $\rightarrow$ | $f_{108}(0)$ | $f_{108}(0)$ | $f_{108}(0)$ | $f_{108}(0)$ | $\rightarrow$ | 0 | 0 | 0 | 0 | 0 |
| 1 | 0 | 0 | 0 | 1 | $\rightarrow$ | $f_{108}(4)$ | $f_{108}(0)$ | $f_{108}(1)$ | $f_{108}(2)$ | $\rightarrow$ | 0 | 0 | 0 | 1 | 1 |
| 2 | 0 | 0 | 1 | 0 | $\rightarrow$ | $f_{108}(0)$ | $f_{108}(1)$ | $f_{108}(2)$ | $f_{108}(4)$ | $\rightarrow$ | 0 | 0 | 1 | 0 | 2 |
| 3 | 0 | 0 | 1 | 1 | $\rightarrow$ | $f_{108}(4)$ | $f_{108}(1)$ | $f_{108}(3)$ | $f_{108}(6)$ | $\rightarrow$ | 0 | 0 | 1 | 1 | 3 |
| 4 | 0 | 1 | 0 | 0 | $\rightarrow$ | $f_{108}(1)$ | $f_{108}(2)$ | $f_{108}(4)$ | $f_{108}(0)$ | $\rightarrow$ | 0 | 1 | 0 | 0 | 4 |
| 5 | 0 | 1 | 0 | 1 | $\rightarrow$ | $f_{108}(5)$ | $f_{108}(2)$ | $f_{108}(5)$ | $f_{108}(2)$ | $\rightarrow$ | 1 | 1 | 1 | 1 | 15 |
| 6 | 0 | 1 | 1 | 0 | $\rightarrow$ | $f_{108}(1)$ | $f_{108}(3)$ | $f_{108}(6)$ | $f_{108}(4)$ | $\rightarrow$ | 0 | 1 | 1 | 0 | 6 |
| 7 | 0 | 1 | 1 | 1 | $\rightarrow$ | $f_{108}(5)$ | $f_{108}(3)$ | $f_{108}(7)$ | $f_{108}(6)$ | $\rightarrow$ | 1 | 1 | 0 | 1 | 13 |
| 8 | 1 | 0 | 0 | 0 | $\rightarrow$ | $f_{108}(2)$ | $f_{108}(4)$ | $f_{108}(0)$ | $f_{108}(1)$ | $\rightarrow$ | 1 | 0 | 0 | 0 | 8 |
| 9 | 1 | 0 | 0 | 1 | $\rightarrow$ | $f_{108}(6)$ | $f_{108}(4)$ | $f_{108}(1)$ | $f_{108}(3)$ | $\rightarrow$ | 1 | 0 | 0 | 1 | 9 |
| 10 | 1 | 0 | 1 | 0 | $\rightarrow$ | $f_{108}(2)$ | $f_{108}(5)$ | $f_{108}(2)$ | $f_{108}(5)$ | $\rightarrow$ | 1 | 1 | 1 | 1 | 15 |
| 11 | 1 | 0 | 1 | 1 | $\rightarrow$ | $f_{108}(6)$ | $f_{108}(5)$ | $f_{108}(3)$ | $f_{108}(7)$ | $\rightarrow$ | 1 | 1 | 1 | 0 | 14 |
| 12 | 1 | 1 | 0 | 0 | $\rightarrow$ | $f_{108}(3)$ | $f_{108}(6)$ | $f_{52}(4)$ | $f_{108}(1)$ | $\rightarrow$ | 1 | 1 | 0 | 0 | 12 |
| 13 | 1 | 1 | 0 | 1 | $\rightarrow$ | $f_{108}(7)$ | $f_{108}(6)$ | $f_{108}(5)$ | $f_{108}(3)$ | $\rightarrow$ | 0 | 1 | 1 | 1 | 7 |
| 14 | 1 | 1 | 1 | 0 | $\rightarrow$ | $f_{108}(3)$ | $f_{108}(7)$ | $f_{108}(6)$ | $f_{108}(5)$ | $\rightarrow$ | 1 | 0 | 1 | 1 | 11 |
| 15 | 1 | 1 | 1 | 1 | $\rightarrow$ | $f_{108}(7)$ | $f_{108}(7)$ | $f_{108}(7)$ | $f_{108}(7)$ | $\rightarrow$ | 0 | 0 | 0 | 0 | 0 |

(b)

[Table-6: (a) shows the functional values obtained using algorithm 3.1 to draw the complete STD for Rule 108 UCA4NB condition and (b) shows the same for PB]

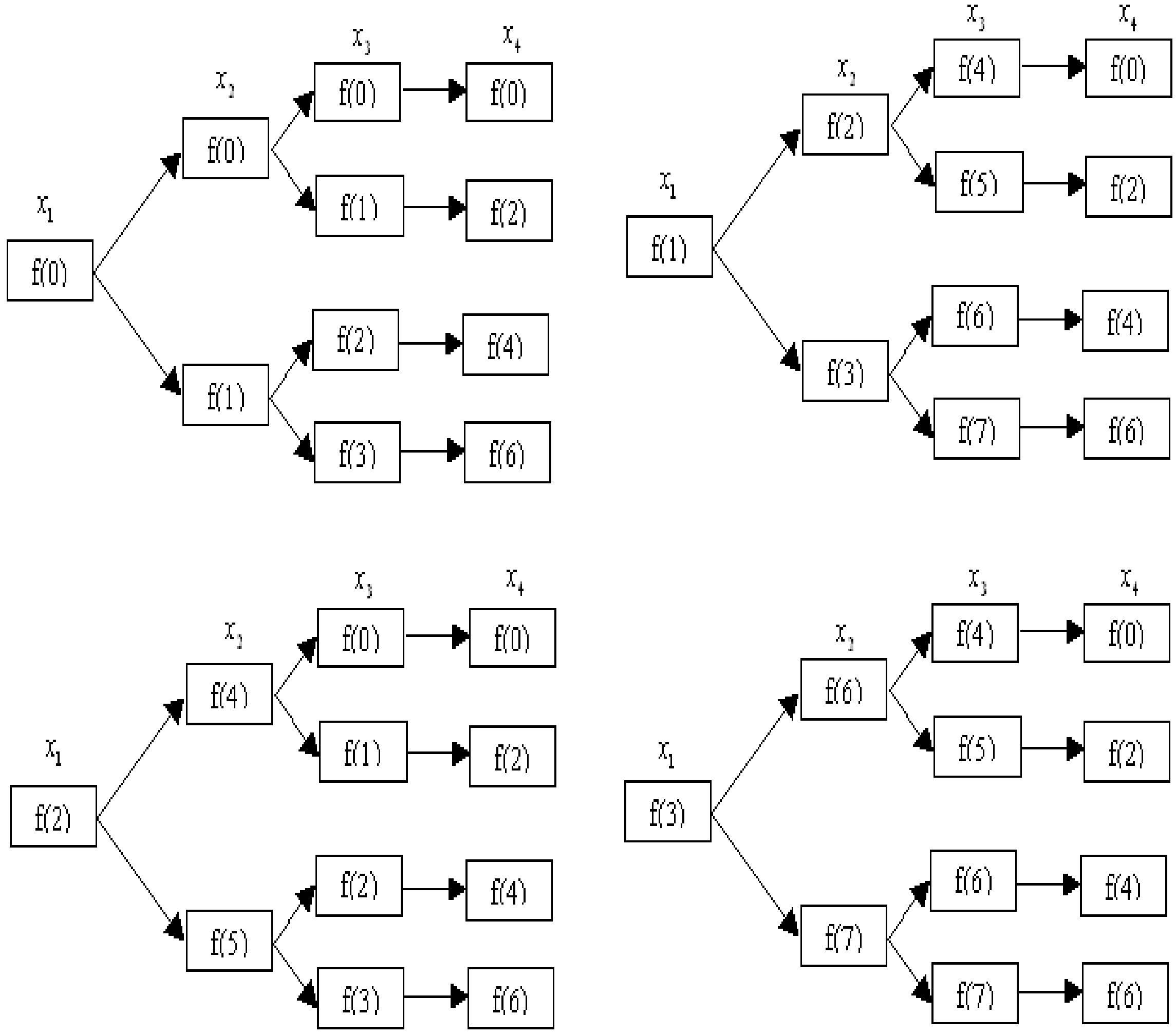

(a)

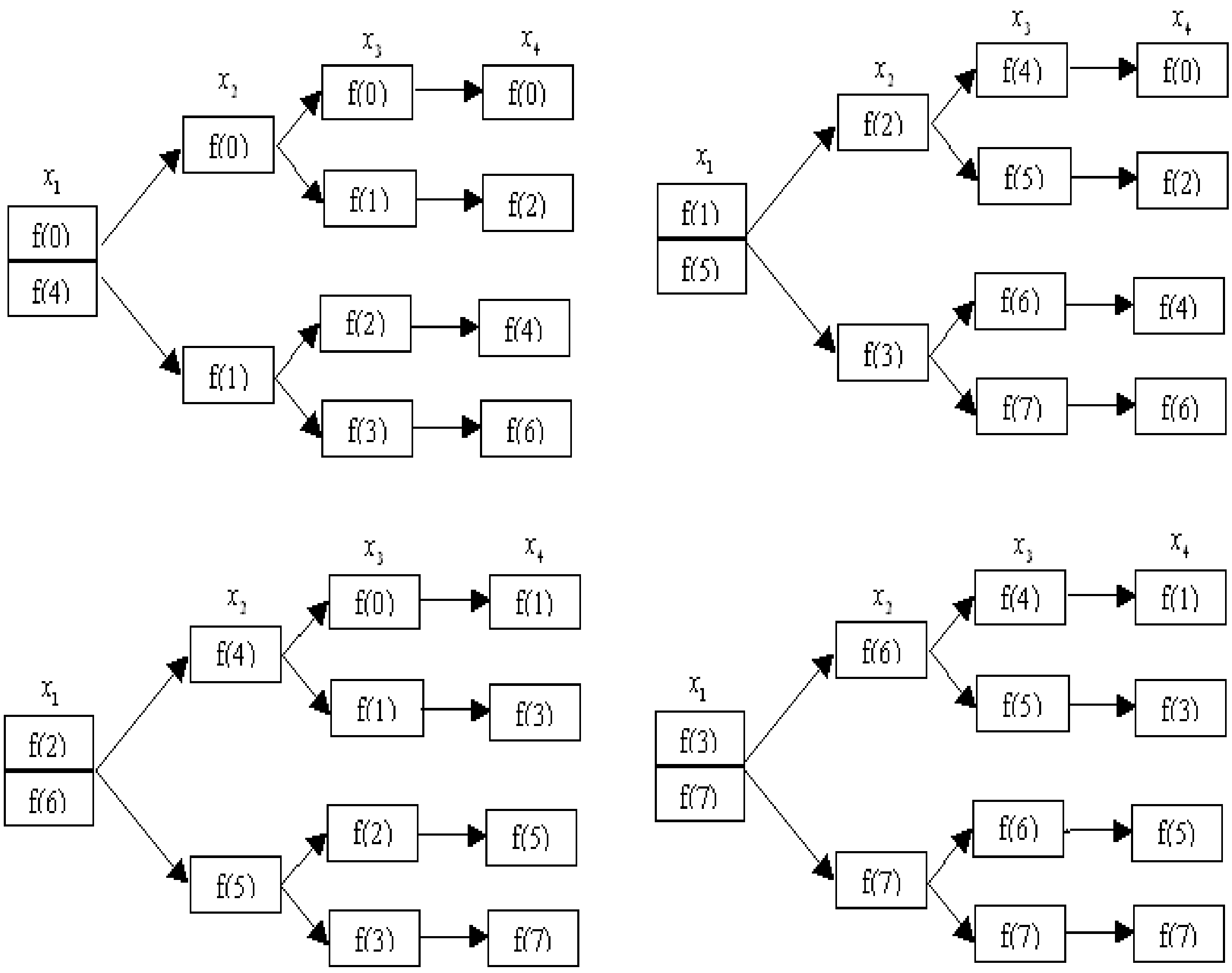

(b)

[Fig-10: (a) shows a binary tree representation for 1D UCA4PB rule 108, obtained using algorithm 3.2. Here $x_i, i = 1, 2 \ldots 5$ denotes the updated value of $x_i$ at time (t+1) after the CA rule $f_{108}$ is applied to it and (b) shows the same for PB condition.]

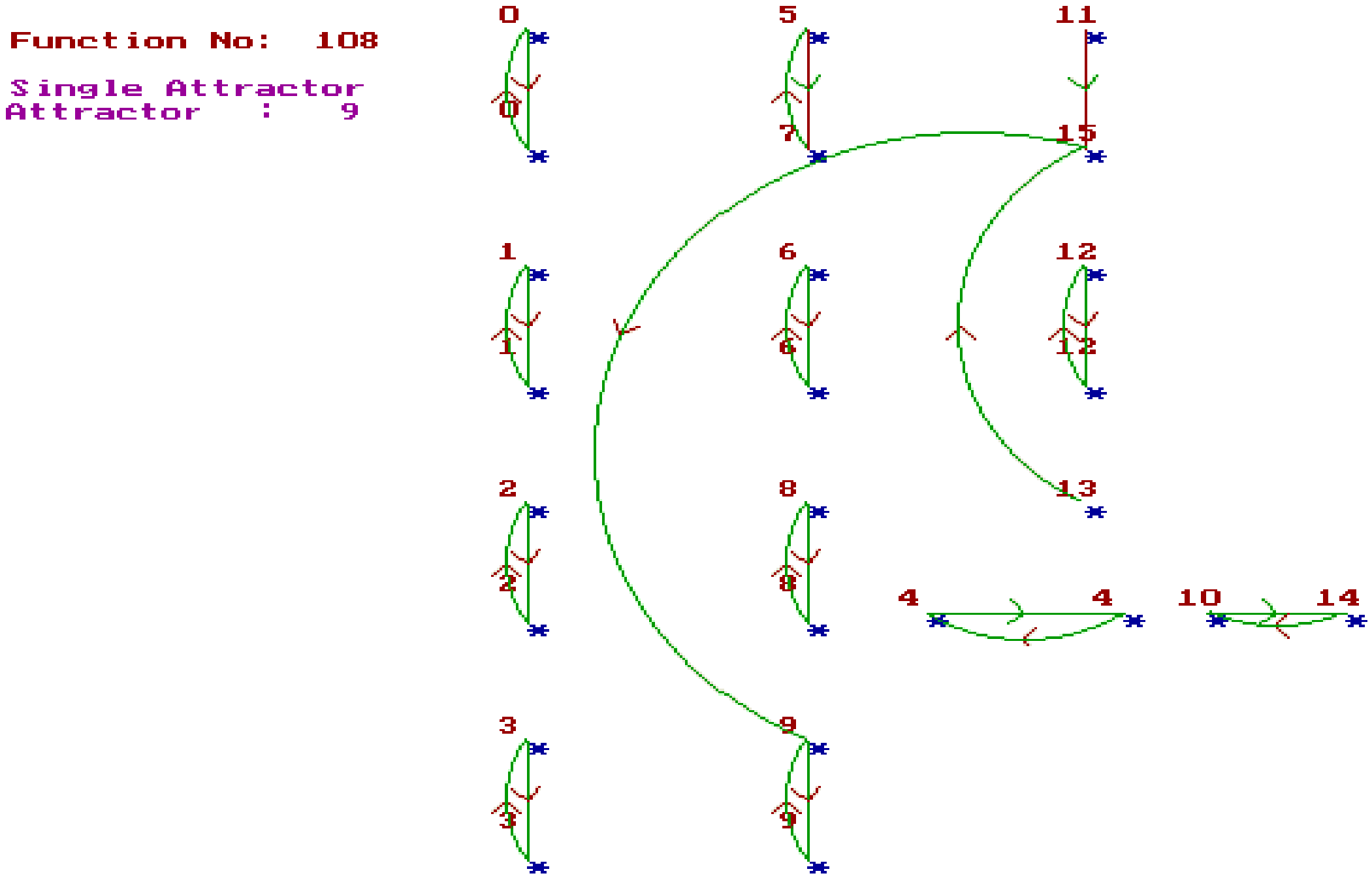

(a)

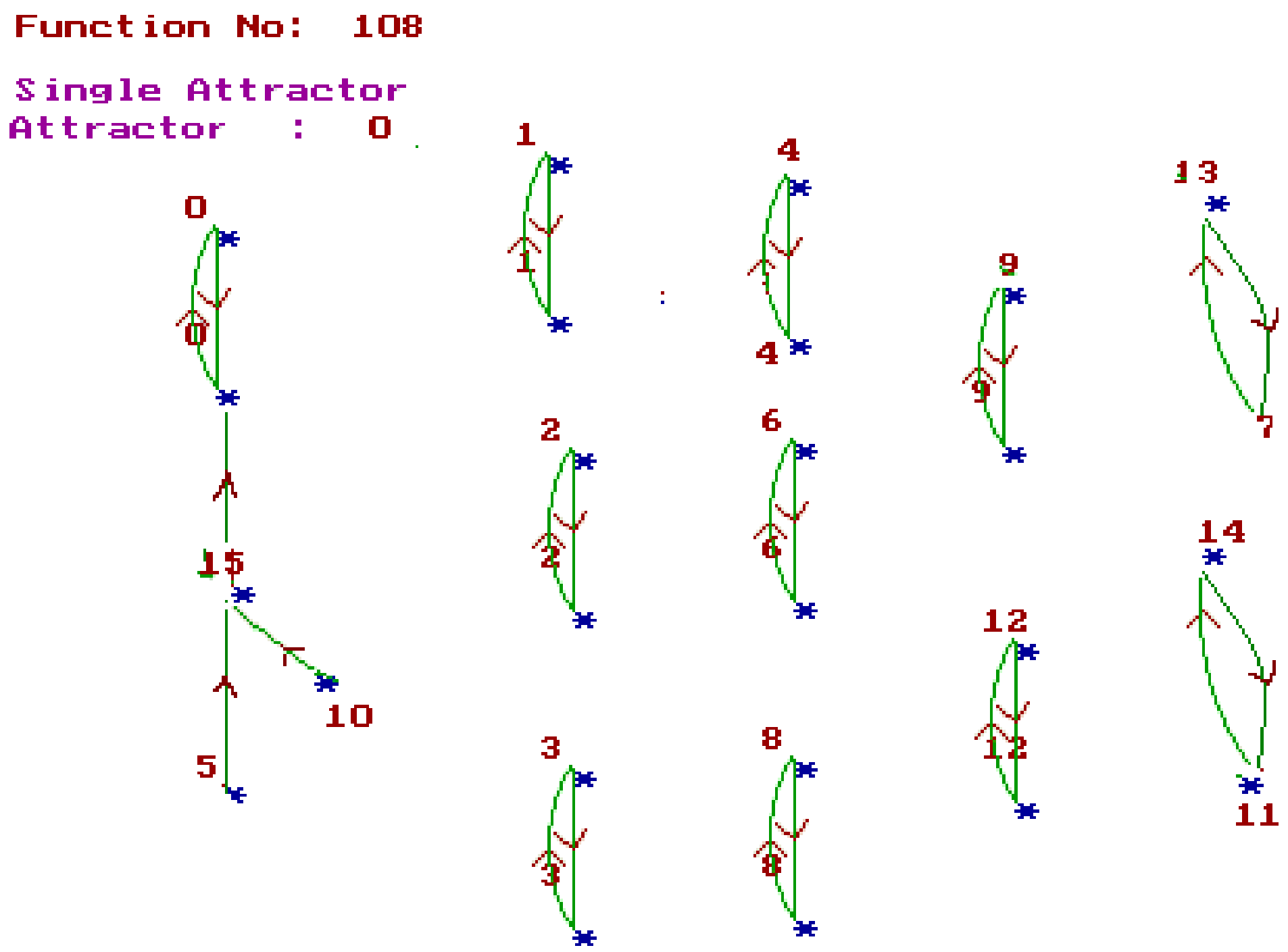

(b)

[Fig-11: (a) shows the complete STD for Rule 108 UCA4NB and (b) shows the same for PB]